\documentclass[10pt,journal]{setup/IEEEtran}
\IEEEoverridecommandlockouts % Comment this out if no funding/thanks.

\makeatletter
\let\NAT@parse\undefined
\makeatother

\usepackage{amsmath,amssymb,amsfonts,amsthm,commath,esint}
\usepackage{bm} % for bold math
\usepackage[numbers,sort&compress]{natbib}

\usepackage{mathtools}
\usepackage{multirow}
\usepackage{physics} % Defines \dif operator, \eval, \norm, \abs and many others
\usepackage{color} % Colour control for LaTeX documents
\usepackage[table]{xcolor} 
\usepackage{makecell}

    \DeclareFontFamily{U}{wncy}{}
    \DeclareFontShape{U}{wncy}{m}{n}{<->wncyr10}{}
    \DeclareSymbolFont{mcy}{U}{wncy}{m}{n}
    \DeclareMathSymbol{\Sh}{\mathord}{mcy}{"58} 

\usepackage{glossaries}  % Create glossaries and lists of acronyms
\setacronymstyle{long-short}

\newacronym{bs}{BS}{base station}
\newacronym{ap}{AP}{access point}
\newacronym{dft}{DFT}{discrete Fourier transform}
\newacronym{dtft}{DTFT}{discrete-time Fourier transform}
\newacronym{ue}{UE}{user equipment}
\newacronym{rf}{RF}{radio-frequency}
\newacronym{los}{LoS}{Line of sight}
\newacronym{ofdm}{OFDM}{orthogonal frequency-division multiplexing}
\newacronym{otfs}{OTFS}{orthogonal time frequency space}
\newacronym{mimo}{MIMO}{multiple-input-multiple-output}
\newacronym{mmimo}{mMIMO}{massive multiple-input-multiple-output}
\newacronym{mmse}{MMSE}{minimum mean squared error}
\newacronym{nmse}{NMSE}{normalized mean squared error}
\newacronym{lmmse}{LMMSE}{linear minimum mean squared error}
\newacronym{mse}{MSE}{mean squared error}
\newacronym{sinr}{SINR}{signal-to-interference-plus-noise ratio}
\newacronym{snr}{SNR}{signal-to-noise ratio}
\newacronym{pec}{PEC}{perfect electric conductor}
\newacronym{ris}{RIS}{reflective intelligent surface}
\newacronym{cms}{CMS}{canonical minimum-scattering}
\newacronym{dma}{DMA}{dynamic metasurface antenna}
\newacronym{elaa}{ELAA}{extremely large aperture antenna}
\newacronym{lis}{LIS}{large intelligent surface}
\newacronym{csi}{CSI}{channel state information}
\newacronym{mf}{MF}{matched filter}
\newacronym{ula}{ULA}{uniform linear array}
\newacronym{upa}{UPA}{uniform planar array}
\newacronym{upw}{UPW}{uniform plane wave}
\newacronym{gstc}{GSTC}{generalized sheet transition condition}
\newacronym{fmcw}{FMCW}{frequency modulated continuous wave}
\newacronym{rcs}{RCS}{radar cross section}
\newacronym{adc}{ADC}{analog-to-digital converter}
\newacronym{dof}{DoF}{degrees of freedom}
\newacronym{papr}{PAPR}{peak to average power ratio}
\newacronym{toa}{ToA}{time of arrival}
\newacronym{jcas}{JCAS}{joint communication and sensing}
\newacronym{cpofdm}{CP-OFDM}{cyclic-prefixed OFDM}

\usepackage{graphicx}  % Enhanced support for graphics
\usepackage{enumitem}  % Control for itemize, numerate, and description
\usepackage{optidef} % For defining optimization problems
\usepackage{mleftright}
\usepackage{cuted}
\usepackage{algorithm2e}
\usepackage{microtype}
\usepackage[binary-units, per-mode=power]{siunitx}
\usepackage{ifthen} % Used for macros
\usepackage{float}

\SetKwInput{kwInit}{Initialization}

\ifCLASSOPTIONcompsoc
    \usepackage[caption=false, font=normalsize, labelfont=sf, textfont=sf]{subfig}
\else
\usepackage[caption=false, font=footnotesize]{subfig}
\fi

\makeatletter
\def\blfootnote{\xdef\@thefnmark{}\@footnotetext}
\makeatother

\newcommand{\iu}{\mathrm{i}\mkern1mu} %ISO 80000-2:2009

\newcommand{\es}[1]{\sin{\left(#1\right)}}
\newcommand{\ec}[1]{\cos{\left(#1\right)}}
\newcommand{\p}[1]{\left(#1\right)}

\renewcommand{\vec}[1]{\mathbf{\lowercase{#1}}}
\newcommand{\bvec}[1]{\bar{\mathbf{\lowercase{#1}}}}
\newcommand{\hvec}[1]{\hat{\mathbf{\lowercase{#1}}}}

\newcommand{\mat}[1]{\mathbf{\uppercase{#1}}}

\newcommand{\mati}[1]{\pmb{\uppercase{#1}}}
\newcommand{\eye}{\mat{I}}
\newcommand{\veci}[1]{\pmb{\lowercase{#1}}}
\newcommand{\bveci}[1]{\bar{\pmb{\lowercase{#1}}}}

\newcommand{\T}{^\mathsf{T}}    % transpose
\newcommand{\conj}{^*}    % transpose
\renewcommand{\H}{^\mathsf{H}}   % hermitian

\newcommand{\e}[1]{\mathrm{e}^{#1}}

\DeclareMathOperator{\diag}{diag}
\newcommand{\hadamard}{\odot} % pointwise multiplication
\DeclareMathOperator{\Imag}{Im}

\newcommand*\Eval[3]{\left.#1\right\rvert_{#2}^{#3}}

\DeclarePairedDelimiterXPP\Aver[1]{\mathbb{E}}{[}{]}{}{

#1
}

\makeatletter
\newcommand*\dotp{\mathpalette\dotp@{.5}}
\newcommand*\dotp@[2]{\mathbin{\vcenter{\hbox{\scalebox{#2}{$\m@th#1\bullet$}}}}}
\makeatother

\newcommand{\der}[2]{\frac{\partial \,#1}{\partial \,#2}}

\definecolor{sepia}{rgb}{0.44, 0.26, 0.08}
\definecolor{pink}{rgb}{1, 0.41, 0.79}
\definecolor{dkgreen}{rgb}{0.4, 0.7, 0.2}

\newcommand{\pre}[1]{{\color{dkgreen}{\textit{\scriptsize PRE: #1}}}}

\newtheorem{remark}{Remark}

\usepackage{verbatim}
\begin{document}

% Title and authors
\title{On Models with Power Conservation in Reflective Intelligent Surfaces and their Design Implications
% Implication of passive lossless metasurface based reconfigurable intelligent surfaces
\thanks{This work was supported by the Villum Investigator Grant
“WATER” from the Velux Foundation, Denmark. The work by P. Ram\'irez-Espinosa has been funded by the European Union under the Marie Sklodowska-Curie grant agreement No. 101109529.}
\thanks{R. J. Williams, O. Semenovska and P. Popovski are with Department of Electronic Systems, Connectivity Section, Aalborg University, 9220 Aalborg \O st, Denmark. E-mail: \{rjw, ose, petarp\}@es.aau.dk.}
\thanks{O. Semenovska is also with Electronic Engineering Department, National Technical University of Ukraine, Ukraine.}
\thanks{P. Ram\'irez-Espinosa is with Telecommunications Research Institute (TELMA), University of M\'alaga,  29071 M\'alaga, Spain. E-mail: pre@ic.uma.es.}
}

\author{Robin J. Williams, Pablo Ram\'irez-Espinosa, Olena Semenovska and Petar Popovski}

\maketitle

\begin{abstract}
Reconfigurable intelligent surfaces (RISs) are potential enablers of future wireless communications and sensing applications and use-cases. The RIS is envisioned as a dynamically controllable surface that is capable of transforming impinging electromagnetic waves in terms of angles and polarization. Many models has been proposed to predict the wave-transformation capabilities of potential RISs, where power conservation is ensured by enforcing that the scattered power equals the power impinging upon the aperture of the RIS, without considering whether the scattered field adds coherently of destructively with the source field. In effect, this means that power is not conserved, as elaborated in this paper. With the goal of investigating the implications of global and local power conservation in RISs, work considers a single-layer metasurface based RIS. A complete end-to-end communications channel is given through polarizability modeling and conditions for power conservation and channel reciprocity are derived. The implications of the power conservation conditions upon the end-to-end communications channel is analyzed.

%The analysis shows that in multipath scenarios, when adhering to power conservation, the RIS-assisted path can not be formulated in the commonly used cascaded channel formulation; $h = h_1 \Phi h_2$. This has direct implications on channel-estimation procedures based upon the cascaded channel formulation. The analysis however shows that given a sufficient number of elements, the error made due to the invalidity of the cascaded channel formulation tends to zero, which implies that conventional channel-estimation procedures can be applied to large RISs with negligible performance degradation. The analysis also demonstrates that it is theoretically possible to do spatial multiplexing through a single RIS element, allowing the system reuse the full RIS aperture to serve multiple communication links simultaneously. Additionally, given orthogonal polarization or directions, the analysis also shows that a \gls{ris} is theoretically capable of creating one-way communication channels through the \gls{ris}, breaking channel reciprocity, and enabling new applications and use-cases.
\end{abstract}
\begin{IEEEkeywords}
Reconfigurable Intelligent Surface (RIS), metasurface, non-reciprocity, power conservation, channel estimation, scattering, radar cross section (RCS), mutual coupling, farfield, polarizability.
\end{IEEEkeywords}
\glsresetall
\section{Introduction}\label{sec:introduction}
%\rjw{Hello}
%\pp{Hello}
%\ose{Hello}
%\kchu{Hello}
%
%
The term \Glspl{ris}, also called Reconfigurable Intelligent surfaces are used to refer to the electromagnetic surfaces that can be dynamically controlled to change the properties of the wireless propagation environment. Ideally, RISs are conceived as semi-passive electromagnetic sheets whose dynamic impact on the propagated waves can be advantageous in multiple ways, such as communication, sensing, positioning~\cite{bjornson_reconfigurable_2022} or even control of electromagnetic exposure~\cite{ZappoRenzo}.
% \glspl{ris} have been considered as a enabling technology for the next generation of wireless systems . The ability to control the propagation environment at will naturally led to an increasing interest in analyzing the potential benefits in terms of communications, localization, sensing and, in general, its impact on the wireless channel---considered thus far as imposed by the propagation environment. 

A model commonly used to capture the impact of RIS on the wireless propagation is the \emph{cascaded channel model}, where a RIS is modeled as a collection of sub-wavelength reflecting elements, resulting in an equivalent channel at the receiver~\cite{yu_review_2023}:
\begin{align}
    \mat{H} = \mat{H}_1 \mat{\Phi} \mat{H}_2, \label{eq:cascadedChannel}
\end{align}
%where $N_\text{rx}$, $N_\text{tx}$, $N$ is the number of receive antennas, transmit antennas, and \gls{ris} elements respectively, $\mat{H}_1 \in \mathbb{C}^{N_\text{rx} \times N}$ is the channel matrix from the transmitter to the RIS, $\mat{H}_2 \in \mathbb{C}^{N \times N_\text{tx}}$ is the channel from the RIS to the receiver, and $\mat{\Phi}\in\mathbb{C}^{N \times N}$ is a reflection matrix modeling the reflection coefficient of each individual element of the RIS. 
where $\mat{H}_1$ is the channel matrix from the transmitter to the \gls{ris}, $\mat{H}_2$ is the channel from the \gls{ris} to the receiver, and $\mat{\Phi}$ is a reflection matrix. In the simplest case, this reflection matrix is a diagonal matrix of unit amplitude entries, representing loss-less reflection of arbitrary phase. Due to its simplicity and the independence between $\mat{\Phi}$ and the different channels, a plethora of works has dealt with channel estimation procedures, beamforming algorithms, RIS-based localization systems, and in general, problems where the target is optimizing the reflection matrix $\mat{\Phi}$ with respect to any performance metric~\cite{yu_review_2023,hassouna_survey_2023}. 

%A new concept is coined in  \cite{shen_modeling_2022}, introducing the term 'beyond-diagonal' RIS where the off-diagonal entries of the reflection matrix $\mat{\Phi}$ can take non-zero values. This is equivalent of a controllable coupling between the RIS elements, as the non-zero off-diagonal elements of the reflection matrix allows a RIS element to reflect the field received by another element. Although mutual coupling is naturally present in RIS, being able to arbitrary control it requires complex and potentially active designs. 

Inheriting this simple cascaded channel, \cite{bjornson_rayleigh_2020,bjornson_power_2020} derived both the spatial correlation of $\mat{H}_1$ and $\mat{H}_2$ under Rayleigh fading and the channel matrices themselves in a line-of-sight deterministic channel based on an effective electrical aperture model of the \gls{ris} elements. A similar approach is used in \cite{tang_wireless_2021}, where each element has a radiation pattern and an aperture, being thus the power reflected by the element given by the product of the radiation pattern and the surface power density of the source field projected onto the aperture of the element. 
%Based on this simple version of the cascaded channel model many channel estimation procedures and and beamformers has been proposed, optimizing the reflection matrix with respect to everything from single user signal-to-noise ratio (SNR), to multi-user minimum rate .  
%As such, extensive research has been done on optimizing RIS-assisted communication systems, given arbitrary realisations of the channel matrices $\mat{H}_1$ and $\mat{H}_2$.
However, as proved in \cite{williams_communication_2020}, the aperture of the reflecting element and its radiation pattern cannot be treated independently without violating power conservation. Central to all these works based on the cascaded channel is the fact that they rely on the strong assumption that the channel can be separated into three independent parts, i.e., $\mat{H}_1$ and $\mat{H}_2$ are completely independent of $\mat{\Phi}$. This assumption, though, has not been proven in general. 

\begin{figure}[t ]
    \centering
    \includegraphics[width=\linewidth]{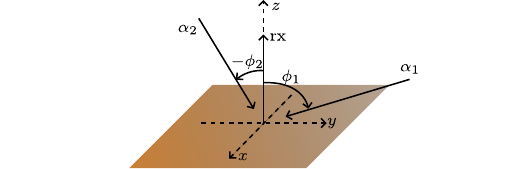}
    \caption{Illustration of a \gls{ris} being illuminated by two planewaves. The impinging waves are scattered towards a user (rx) located in the broadside direction.}
    \label{fig:intro_scenario}
\end{figure}
As an illustrative example of the present work, consider the scenario in Fig. \ref{fig:intro_scenario} with a single electrically small \gls{ris} element that is illuminated by two plane waves with amplitudes $\alpha_1$ and $\alpha_2$ from two directions $\phi_1$ and $\phi_2$. The element scatters the impinging waves towards a single receiver located in the $\phi_\text{out} = 0$ direction. The incoming and outgoing waves have the same polarization, and the element is constructed such that it acts as a reflector when illuminated from the direction $\phi = 0$. As we shall explain later on in Section \ref{sec:impact_commun}, the \gls{ris}-assisted channel is given as $h = C (g_1+g_2)$ where
\begin{align}
    %h &= \frac{ 3 \eta }{16 \pi}  \frac{\e{-\iu k x_\text{rs}}}{ x_\text{rs}}   a_\text{r}  \p{g_1 + g_2}  \\
    g_n &= \alpha_n \p{\e{\iu \rho} \p{1+\ec{\phi_n}} - \iu \p{1-\ec{\phi_n}}  }
\end{align}
and $\rho$ is the phase-shift introduced by the element, with $C$ and $\alpha_n$ some constants. From this simple example, we see that if and only if $\phi_n = 2 m \pi$ for $m$ a natural number, then  $g_n = \alpha_n \e{\iu \rho}$ such that the channel admits the cascaded expression
%\begin{align}
 %   h =\underbrace{\frac{ 3 \eta }{16 \pi}  \frac{\e{-\iu k x_\text{rs}}}{ x_\text{rs}}   a_\text{r}}_{h_1} \underbrace{\vphantom{\frac{\e{-\iu k x_\text{rs}}}{ x_\text{rs}}}   \e{\iu \rho}}_{\Phi} \underbrace{\vphantom{\frac{\e{-\iu k x_\text{rs}}}{ x_\text{rs}}}\p{\alpha_1 + \alpha_2}}_{h_2}.
%\end{align}
\begin{align}
    h =\underbrace{C}_{h_1} \underbrace{\e{\iu \rho}}_{\Phi} \underbrace{\p{\alpha_1 + \alpha_2}}_{h_2}.
\end{align}
In other words, the cascaded channel is valid \textit{only} for a single spatial direction.

%\pp{PP: My understanding is to use this simple example and show that the cascaded model and the other prior model violate power conservation? Basically, telling why the phase and the amplitude cannot be manipulated independently.}

%A partial answer is provided in \cite{gradoni_end--end_2020}, showing that the cascaded channel model holds if power conservation is ensured, although at the price of a significant reduction in the reflection phase-coverage. Along this line, this work shows that by including magnetic moments in the analysis (and hence capturing the behavior of a general reflecting element), the phase-coverage can be extended, but at the cost of independence between the three channel parts, i.e., the validity of the cascaded channel.

%The limitations of the cascaded channel to capture the electromagnetic phenomena inherent to RIS have been, up to some extent, unveiled, and alternative models have already been considered. 
Being aware of the limitations of the cascaded channel, more sophisticated models have been proposed, aiming to capture the electromagnetic phenomena inherent to \glspl{ris} and rather neglected in the cascaded model. In \cite{di_renzo_communication_2022}, the RIS is modeled as an in-homogeneous impedance boundary subject to the physical optics approximation. The result is a model that is coherent with Maxwell's equations, and links the surface impedance to the reflection properties of the surface together with a set of conditions for power conservation. However, the model does not allow enforcing of power conservation on a microscopic scale, nor does it allow polarization transformations or breaking of reciprocity, as demonstrated using metasurface based designs \cite{achouri_fundamental_2021,achouri_electromagnetic_2021}. 

Approaches based on impedance models are presented in \cite{gradoni_end--end_2020, Konno2024}, where the response of the \gls{ris} is no longer a linear term as in \eqref{eq:cascadedChannel}. They show, though, that this modified cascaded channel is valid if power conservation is ensured, although at the price of a significant reduction in the reflection phase-coverage. In fact, the elements phase-shift coverage is restricted to $\frac{\pi}{2}$ for a reflection coefficient amplitude of above $0.75$, which is a consequence of considering a single electric dipole moment. Despite the limitations, the models are, nevertheless, useful to capture some physical limitations of \glspl{ris}, and its application to communications have been studied in  \cite{qian_mutual_2021,abrardo_mimo_2021,hassani_optimization_2023}.

Scattering parameters-based formulations are given in \cite{shen_modeling_2022,abeywickrama_intelligent_2020}. The former also assumes the off-diagonal entries of the reflection matrix $\mat{\Phi}$ can take non-zero values. This is equivalent to a controllable mutual coupling, which would require complex and potentially active \gls{ris} designs. As is shown in \cite{nossek_physically_2023}, modeling of scattering is equivalent to modeling impedance parameters. By accounting for the fact that the total current flowing on the reflector is the superposition of both the current induced by the impinging wave and the currents reflected by an attached load, as opposed to only the reflected currents, the restricted phase-response remains. The work \cite{nerini_universal_2023} brings further analytical details about the equivalence of scattering parameter, impedance parameter, and admittance parameter modeling, while \cite{nerini_physically_2024} extends the modeling to a layered architecture of surfaces capable of spatial domain processing. 

%A different approach based on mutual impedances \cite{ivrlac_toward_2010} was presented in \cite{gradoni_end--end_2020}. The model accounts for an induced electric dipole moment along a single principal axis. The result of this is that the elements phase-shift coverage is restricted to $\frac{\pi}{2}$ for a reflection coefficient amplitude of above $0.75$. As a result, the surface is only purely reflective for a specific reflection phase. As the phase is changed, the surface becomes increasingly transmissive which is not something that is observed in practical implementations \cite{tang_wireless_2021}. The modelling approach however ensures that power is correctly conserved at a microscopic scale. The impedance model is, nevertheless, useful to capture some physical limitations of RIS, and its application to communications have been studied in  \cite{qian_mutual_2021,abrardo_mimo_2021,hassani_optimization_2023}.

Despite the limitations shown in the previous works, it is known that metasurface based designs can provide full $2\pi$ phase-coverage in the reflection coefficient \cite{achouri_electromagnetic_2021}. Additionally, new and exotic wave interactions such as polarization transformations and breaking of channel reciprocity have been demonstrated in \cite{achouri_general_2015,achouri_fundamental_2021}. The purpose of this paper is to introduce advances in metasurface modelling to the wireless community and thereby find potential applications and new research directions utilizing the exotic wave interactions enabled by metasurface based designs. Specifically, our contributions are
\begin{itemize}
    \item We model the \gls{ris} as a collection of electrically small particles with dipolar response, including electric and magnetic moments, and characterized through their \textit{polarizability} matrices. We clearly show the conditions under which these matrices represent \textit{passive} and \textit{reciprocal} \glspl{ris}, both widely assumed properties.
    \item We demonstrate that the particles can be designed to modify phase, amplitude and polarization of the impinging wave, and detail the relationship between these magnitudes and the angular directions, proving that phase and amplitude are correlated, and showing when the cascaded channel becomes valid. 
    \item The consequences of these findings in wireless communications are analyzed, questioning the standard channel estimation methods and characterizing the asymptotic behaviour of the \gls{ris} (large number of elements).
    \item We present a general framework to optimize the \gls{ris} aiming to maximize its utility in wireless, and show results where it is theoretically possible to improve the performance (e.g., phase coverage) compared to state-of-the-art models. 
\end{itemize}

%\textbf{We want to highlight the difference to other works. The goal of this work is to show what is theoretically possible when assuming a very simple structure to provide some improved degree of service compared to conventional models based upon conventional antenna designs.} \pp{PP: This part describes the paper contribution, needs to be expanded and strengthened.}

\textit{Paper overview:} 
Sec. \ref{sec:systemModel} starts with a short review of the prerequisite electromagnetic theory and presents the end-to-end channel model. Sec. \ref{sec:properties} analysis the relationship between phase, amplitude, polarization and angular direction for  passive and lossless \glspl{ris}. Sec. \ref{sec:impact_commun} analyzes the impact of such trade-off in common communications scenarios. Sec. \ref{sec:optimization} treats the optimization of the \gls{ris} configuration with respect to the received signal power. Sec. \ref{sec:numericalResults} presents a range of simulation results and, finally, Sec. \ref{sec:conclusion} concludes the work and highlights main results.

\textit{Notation:} $\iu$ is the imaginary unit, $\norm{\cdot}_2$ is the Euclidean norm, $\abs{\cdot}$ is the absolute value, $\cdot^T$ and $\cdot^H$ are the transpose and Hermitian transpose respectively. Vectors are denoted by bold lowercase symbols, and matrices are denoted by bold uppercase symbols. $\p{\mat{A}}_{n,m}$ denotes the element on the $n$'th row and $m$'th column of $\mat{A}$. $\Tr{\cdot}$ is the trace operator and $\Aver{x}$ is the expectation operator. $\diag\p{\mat{A}}$ is a vector made the entries along the diagonal of $\mat{A}$ and $\hadamard$ is the Hadamard product. The shorthand notation $r = \norm{\vec{r}}_2$ and $\hvec{r} = \vec{r} r^{-1}$ is used throughout the paper.

%Introduce prior works, specificially
%\begin{enumerate}
%    \item Performance Analysis of Systems with Coupled and Decoupled RISs
%    \item End-to-End Mutual-Coupling-Aware Communication Model for Reconfigurable Intelligent Surfaces: An Electromagnetic Compliant Approach Based on Mutual Impedances
%\end{enumerate}

\section{End-to-end communications channel}\label{sec:systemModel}
% Old section backup:
% \input{systemModel/index.tex}
%
% All the backup files are in systemModel folder

We consider a narrowband \gls{ris}-assisted communication system consisting of one transmitter, one receiver, and one \gls{ris}, all suspended in a vacuum. Both the transmitter and the receiver are single antenna devices located at arbitrary positions $\vec{x}_\text{t}\in\mathbb{R}^{3\times 1}$ and $\vec{x}_\text{r}\in\mathbb{R}^{3 \times 1}$. The \gls{ris} is modeled as a planar antenna structure centered at the origin of coordinates, and contained in the $z=0$ plane (without any loss of generality). The \gls{ris} is made up of electrically small reconfigurable scattering structures, henceforth called particles, similarly as how metasurfaces are constructed. Besides, we assume the thickness of the surface is negligible, following the envisioned idea of passive, cheap and thin \glspl{ris}\footnote{The proposed modeling approach can, however, be easily extended to multiple layers of simple lattice structures, commonly referred to as stacked \glspl{ris} \cite{Jiancheng2023}.}

As an electromagnetic wave from an external source---either the transmitter or any scatterer---hits the \gls{ris}, currents are induced in the particles. The particles scatters a so-called \textit{interaction field} among them which in turn also induces new currents in the particles. This effect creates a dampened oscillation that continues until equilibrium, point at which the \gls{ris} carries statically oscillating electric and effective magnetic current distributions. Depending on these currents and the particle configuration, the impinging field is transmitted or reflected in a different way.

To establish the end-to-end communication channel between the transmitter and the user through both the direct and the \gls{ris}-assisted paths, we first review some fundamentals of electromagnetic radiation.

%---------------------------------------------------------------
% ELECTROMAGNETIC RADIATION
%---------------------------------------------------------------
\subsection{Electromagnetic radiation}

Given some current distribution in space, represented by the electric $\veci{j}\in\mathbb{C}^{3\times 1}$ and magnetic $\veci{m}\in\mathbb{C}^{3\times 1}$ current vectors, the radiated electric $\vec{e}\in\mathbb{C}^{3\times 1}$ and magnetic $\vec{h}\in\mathbb{C}^{3\times 1}$  fields at an arbitrary point are given by \cite[sec. 3.4]{balanis_antenna_2016}
\iffalse
\begin{align}
    \begin{bmatrix}
        \vec{e}\p{\vec{x}} \\ \vec{h}\p{\vec{x}}
    \end{bmatrix} = \iiint_{-\infty}^{\infty} \mat{G}\p{\vec{x} - \bvec{x}} \begin{bmatrix}
        \veci{j}\p{\bvec{x}} \\ \veci{m}\p{\bvec{x}}
    \end{bmatrix} \dd{\bvec{x}}, \label{eq:greensFormulation}
\end{align}
\fi
\begin{align}
    \begin{bmatrix}
        \vec{e}\p{\vec{x} +  \veci{x}} \\ \vec{h}\p{\vec{x} +  \veci{x}}
    \end{bmatrix} = \iiint_{-\infty}^{\infty} \mat{G}\p{\vec{d}  +  \veci{x} - \bveci{x}} \begin{bmatrix}
        \veci{j}\p{\bvec{x} + \bveci{x}} \\ \veci{m}\p{\bvec{x} + \bveci{x}}
    \end{bmatrix} \dd{\bveci{x}}, \label{eq:greensFormulation}
\end{align}
where $\vec{x}\in\mathbb{R}^{3\times 1}$ and $\bvec{x}\in\mathbb{R}^{3\times 1}$ point to the centers of the measurement and source volumes, $\vec{d} = \vec{x} - \bvec{x}$ is the distance vector from the source volume to the measurement volume, $\veci{x}\in\mathbb{R}^{3\times 1}$ and $\bveci{x}\in\mathbb{R}^{3\times 1}$ indicate a relative position w.r.t. $\vec{x}$ and $\bvec{x}$, and $\mat{G}(\cdot)\in\mathbb{C}^{6\times 6}$ is the free-space Green's function.   

The Green's matrix is often written out in terms of the sub-matrices $\mat{G}_\text{ee}$, $\mat{G}_\text{em}$, $\mat{G}_\text{me}$, and $\mat{G}_\text{mm}$, denoting the electric-to-electric, magnetic-to-electric, electric-to-magnetic, and magnetic-to-magnetic spatial impulse-responses, respectively. Thus, we have 
\begin{align}
        \mat{G}\p{\vec{r}} =& \iu k \begin{bmatrix}
        \mat{G}_\text{ee}\p{\vec{r}} \eta & \mat{G}_\text{em}\p{\vec{r}} \\
        \mat{G}_\text{me}\p{\vec{r}}& \mat{G}_\text{mm}\p{\vec{r}} \eta^{-1},
    \end{bmatrix} \label{eq:greensMatrix}
\end{align}
where $k = 2\pi/\lambda$ is the wavenumber, $\eta = \sqrt{\mu\epsilon^{-1}}$ is the free space impedance ($\mu$ and $\epsilon$ are the permeability and permittivity), and the distinct submatrices are given by
\begin{align}
    \mat{G}_\text{ee}\p{\vec{r}} = \mat{G}_\text{mm}\p{\vec{r}} =&\frac{-1}{k^2} \p{k^2 \eye_3 + \nabla \nabla\T} \frac{\e{-\iu k r}}{4 \pi r}, \label{eq:Gee}\\
    \mat{G}_\text{em}\p{\vec{r}} = -\mat{G}_\text{me}\p{\vec{r}} =& \frac{-1}{\iu k}   \p{\nabla \cross \frac{\e{-\iu k r}}{4 \pi r}}. \label{eq:Gem}
\end{align}

If $\vec{r}\rightarrow \vec{0}$, then the following relation holds:
\begin{align}
\mat{G}_0&=\lim_{\vec{r} \to \vec{0}_{3\times 1}} \Re{\mat{G}\p{\vec{r}}} = \frac{-k^2}{6 \pi} \begin{bmatrix}
  \eye_3 \eta & \mat{0}_{3\times 3} \\
\mat{0}_{3\times 3} & \eye_3 \eta^{-1}
\end{bmatrix}. \label{eq:G0}
\end{align}
In turn, in the far field ($r$ is large) the higher order terms of the Green's matrices vanish, and they are approximated by 
\begin{align}
\begin{split}
    \mat{G}_\text{ee}^{\text{ff}}&\p{\vec{d} +  \veci{x} - \bveci{x} } = \frac{\e{-\iu k \p{d + \hvec{d}\T\p{\veci{x}-\bveci{x}}} }}{4 \pi d} \mat{P}_1\p{\vec{d}}, \\
    \mat{G}_\text{em}^{\text{ff}}&\p{\vec{d } +  \veci{x} - \bveci{x} } = \frac{\e{-\iu k \p{d + \hvec{d}\T\p{\veci{x}-\bveci{x}}} }}{4 \pi d} \mat{P}_2\p{\vec{d}}, \\
    %\mat{P}_1 &=\frac{1}{d^2} \begin{bmatrix}
   %\p{\vec{d}}_1^2-d^2  &\p{\vec{d}}_1\p{\vec{d}}_2 &\p{\vec{d}}_1\p{\vec{d}}_3 \\
   %\p{\vec{d}}_2\p{\vec{d}}_1 &\p{\vec{d}}_2^2 -d^2&\p{\vec{d}}_2\p{\vec{d}}_3 \\
   %\p{\vec{d}}_3\p{\vec{d}}_1 & \p{\vec{d}}_3\p{\vec{d}}_2 %&\p{\vec{d}}_3^2-d^2
    %\end{bmatrix}, 
    \mat{P}_1\p{\vec{d}} &= d^{-2} \vec{d}\vec{d}^T - \mat{I}_3,\\
    \mat{P}_2\p{\vec{d}} &= \frac{1}{d} \begin{bmatrix}
    0  & -\p{\vec{d}}_3 & \p{\vec{d}}_2 \\
   \p{\vec{d}}_3 & 0 & -\p{\vec{d}}_1 \\
    -\p{\vec{d}}_2 & \p{\vec{d}}_1 & 0
    \end{bmatrix}. 
\end{split} \label{eq:radiativeGreen}
\end{align}
Note that the far field assumption requires $\vec{d}$ large and $\norm{\bveci{x}}_2, \norm{\veci{x}}_2 \ll d$. Inserting \eqref{eq:radiativeGreen} into \eqref{eq:greensFormulation} yields the electromagnetic radiation in the far field as
\begin{align}
    \begin{bmatrix}
        \vec{e}\p{\vec{x} + \veci{x}} \\  \eta \vec{h}\p{\vec{x} + \veci{x}} 
    \end{bmatrix} &=  \frac{\e{-\iu k \p{d + \hvec{d}\T\veci{x} }}}{4\pi d} \begin{bmatrix}
        \vec{p} \\
        \hvec{d} \cross \vec{p}
    \end{bmatrix} \label{eq:farfieldPlanewave}
\end{align}
where
\begin{align}
    \vec{p} &=\iu k\mat{P} \begin{bmatrix}
    \eye_3 \eta & \mat{0}_{3\times 3} \\
     \mat{0}_{3\times 3} & \eye_3 
    \end{bmatrix}\iiint_{\mathcal{V}_\text{s} - \bvec{x}} \e{\iu k \hvec{d}\T\bveci{x}} \begin{bmatrix}
        \veci{j}\p{\bveci{x}} \\  \veci{m}\p{\bveci{x}}
    \end{bmatrix} \dd{\bveci{x}}, \\
\mat{P} &= \begin{bmatrix}
\mat{P}_1\p{\vec{d}} & \mat{P}_2\p{\vec{d}}\end{bmatrix} ,
\end{align}
with $\mathcal{V}_\text{s}$ the source volume. A closer look to \eqref{eq:farfieldPlanewave} reveals that the field observed under far field conditions is described by a plane wave.

%---------------------------------------------------------------
% RIS RADAR CROSS SECTION
%---------------------------------------------------------------
%\subsection{Optimization considerations and radar cross section}
\subsection{End-to-end channel and RIS radar cross section}
\label{subsec:RIS_RCS}

As shown in \eqref{eq:farfieldPlanewave}, the far field wave always takes the shape of a plane wave, whether the wave originates directly from a transmit antenna, or is the result of scattering by the \gls{ris}. Hence we write the \textit{planewave spectrum} of the field radiated by the transmitter as\footnote{Note that we deliberately focus only on the electric field, as for a plane wave $\vec{h}(\vec{x}) = \eta^{-1}\hvec{d}\cross \vec{e}(\vec{x})$.}
\begin{align}
    \veci{e}_\text{t}(\hvec{r}) =  \frac{\eta}{\lambda} \vec{l}_\text{t}\p{\hvec{r}} s \label{eq:et_tx}
\end{align} %\frac{\e{-\iu k \norm{\vec{x}-\vec{x}_\text{t}}_2}}{\norm{\vec{x}-\vec{x}_\text{t}}_2}
where $s\in\mathbb{C}$ is a transmit symbol, $\hvec{r} \in \mathbb{R}^{3\times 1}$ is a direction vector, and $\vec{l}_\text{t}\p{\hvec{r}}$ is a power normalized effective length defined as
\begin{align}
    \vec{l}_\text{t}\p{\hvec{r}} = \frac{\iu}{2 \sqrt{P_\text{t}}}\mat{P}\p{\hvec{r}} \iiint_{\mathcal{V}_\text{t} - \vec{x}_\text{t}} \e{\iu k \hvec{r}\T\veci{x}_\text{t}} \begin{bmatrix}
        \veci{j}\p{\veci{x}_\text{t}} \\ \eta^{-1} \veci{m}\p{\veci{x}_\text{t}}
    \end{bmatrix} \dd{\veci{x}_\text{t}},
\end{align}
where $\mathcal{V}_\text{t}$ is the volume containing the transmitter current distribution, and $P_\text{t}\in\mathbb{R}^+$ is the reference transmitted power, i.e., the transmitted power when $\mathbb{E}[|s|^2] = 1$. From the plane wave spectrum in \eqref{eq:et_tx}, the electric field at point $\vec{x}$ in free space is directly given by $\vec{e}_\text{t}\p{\vec{x}} =\veci{e}_\text{t}\p{\frac{\vec{x}-\vec{x}_\text{t}}{\norm{\vec{x}-\vec{x}_\text{t}}_2}} \frac{\e{-\iu k \norm{\vec{x}-\vec{x}_\text{t}}_2}}{\norm{\vec{x}-\vec{x}_\text{t}}_2}$. 

As the wave in \eqref{eq:et_tx} propagates, it interacts with different scatterers in the environment, including the \gls{ris}. Thus, the wave impinges upon the received through two channels: the direct channel $\vec{h}_\text{d}$ and the \gls{ris}-assisted channel $\vec{h}_\text{d}$; leading the planewave spectrum of the field at the receiver $\veci{e}_\text{r}\p{\hvec{r}}$ given by
\begin{align}
    \veci{e}_\text{r}\p{\hvec{r}} =& \frac{\eta}{\lambda}  \p{\vec{h}_\text{d}\p{\hvec{r}} + \vec{h}_\text{a}\p{\hvec{r}}} s, \\
    \vec{h}_\text{d}\p{\hvec{r}} =& \iint_{\mathcal{S}_\text{t}}  \mat{H}_\text{rt}\p{\hvec{r}, \hvec{r}_\text{t}} \vec{l}_\text{t}\p{\hvec{r}_\text{t}}  \dd{\hvec{r}_\text{t}}, \\
    \vec{h}_\text{a}\p{\hvec{r}} =& \idotsint_{\mathcal{S}_\text{out}, \mathcal{S}_\text{in}, \mathcal{S}_\text{t}} \mat{H}_\text{rs}\p{\hvec{r}, \hvec{r}_\text{out}} \mat{s}\p{\hvec{r}_\text{out}, \hvec{r}_\text{in}} \mat{H}_\text{st}\p{\hvec{r}_\text{in}, \hvec{r}_\text{t}} \notag \\
    &\quad\quad\times\vec{l}_\text{t}\p{\hvec{r}_\text{t}} \dd{\hvec{r}_\text{t}} \dd{\hvec{r}_\text{in}} \dd{\hvec{r}_\text{out}}, \label{eq:ha_vector}
\end{align}
where $\mathcal{S}_\xi \,\forall\, \xi\in\{\text{r}, \text{t}, \text{in}, \text{out}\}$ are unit-spheres and $\mat{H}_\text{xy}$ is the double-directional channel \cite[sec. 6.7]{molisch_wireless_2011} from actor y to x, encapsulating losses, polarization rotations and reflections ($\text{r}$ denotes receiver, $\text{s}$ is the \gls{ris}, and $\text{t}$ is the transmitter). The scattering matrix $\mat{S}$ in \eqref{eq:ha_vector} captures the response of the \gls{ris} as a whole, describing how a impinging wave is transformed by the \gls{ris} in terms of amplitude, phase, and polarization. In other words, the planewave spectrum radiated by the \gls{ris} is described by
\begin{align}
    \veci{e}_\text{s}(\hvec{r}) = \mat{S}\p{\hvec{r}, \hvec{r}_\text{in}} \veci{e}_\text{in}(\hvec{r}_\text{in}) ,\label{eq:eOut}
\end{align}%\frac{\e{-\iu k x}}{x}
where $\veci{e}_\text{in}\p{\hvec{r}_\text{in}} = \iint_{\mathcal{S}_\text{t}}  \mat{H}_\text{st}\p{\hvec{r}_\text{in}, \hvec{r}_\text{t}} \vec{l}_\text{t}\p{\hvec{r}_\text{t}}  \dd{\hvec{r}_\text{t}}$ is the planewave spectrum of the ingoing electric field at the center of the \gls{ris}. 

Assuming the receiver has a power normalized effective length $\vec{l}_\text{r}$, the received signal (with unit $\sqrt{\text{W}}$) is given by
\begin{align}
    v &= \frac{\eta}{2\lambda}\iint_{\mathcal{S}_\text{r}} \vec{l}\H_\text{r}\p{-\hvec{r}_\text{r}} \p{\vec{h}_\text{d}\p{\hvec{r}_\text{r}} + \vec{h}_\text{a}\p{\hvec{r}_\text{r}}} s \dd{\hvec{r}_\text{r}} \notag \\
    &=  \frac{\eta}{2\lambda} \p{h_\text{d}+ h_\text{a}} s \label{eq:recSignal}
\end{align}
where the $2^{-1}$ factor accounts for the power-splitting between the antenna and a matched load ($\sqrt{2^{-1}}$) and the conversion from peak-values to time-averaged values ($\sqrt{2^{-1}}$). Note that \eqref{eq:recSignal} has the form of the conventional signal model assumed in communications, representing hence the end-to-end channel. 

From \eqref{eq:recSignal}, the channel gain is easily computed as 
\begin{align}
   g = \frac{P_\text{rx}}{P_\text{tx}} &= \frac{\abs{v}^2}{\abs{s}^2} = \p{\frac{\eta}{2 \lambda}}^2 \abs{h_\text{d} + h_\text{a}}^2
\end{align}
and, in general, the objective is maximizing $g$ by adjusting the configuration of the \gls{ris}, i.e. $\mat{S}$, to maximize the amplitude of $h_\text{a}$ and align the phases of $h_\text{d}$ and $h_\text{a}$ to ensure coherent summation. To get further insight, we particularize for a line-of-sight channel; that is, a single plane wave. In this case, the channel matrices are described by
\begin{align}
    \mat{H}_\text{xy}\p{\hvec{r}_1, \hvec{r}_2} = \delta\p{\hvec{r}_1 - \hvec{d}_\text{xy}}  \delta\p{\hvec{r}_2 - \hvec{d}_\text{xy}} \eye_3 \frac{\e{-\iu k d_\text{xy}}}{ d_\text{xy}}, 
\end{align}
where $\vec{d}_{xy} = \vec{x}_\text{x} - \vec{x}_\text{y}$ is the distance vector from agent y to agent x. Therefore, the direct and \gls{ris}-assisted complex gains become
\begin{align}
    h_\text{d} &= \frac{\e{-\iu k d_\text{rt}}}{ d_\text{rt}}  \vec{l}\H_\text{r}\p{-\hvec{d}_\text{rt}} \vec{l}_\text{t}\p{\hvec{d}_\text{rt}}, \\
    h_\text{a} &= \frac{\e{-\iu k (d_\text{rs} + d_\text{st})}}{ d_\text{rs} d_\text{st}} \vec{l}\H_\text{r}\p{-\hvec{d}_\text{rt}} \mat{S}\p{\hvec{d}_\text{rt}, \hvec{d}_\text{st}} \vec{l}_\text{t}\p{\hvec{d}_\text{st}} ,\label{eq:LOSChannel}
\end{align}
and maximizing $g$ becomes equivalent to maximizing
\begin{align}
   f = \big|a_\text{d} \hvec{l}&\H_\text{r}\p{-\hvec{d}_\text{rt}} \hvec{l}_\text{t}\p{\hvec{d}_\text{rt}} \e{-\iu k d_\text{rt}} \notag \\
   &+ a_\text{a} \hvec{l}\H_\text{r}\p{-\hvec{d}_\text{rs}}  \hvec{p}_\text{r} \p{\hvec{d}_\text{rs}, \hvec{d}_\text{st}} \e{-\iu k \p{d_\text{rs} + d_\text{st}}}  \big|^2,\label{eq:optimizationObjectiveF}
\end{align}
where 
\begin{align}
    a_\text{d} =&  d_\text{rt}^{-1}\sqrt{G_\text{r}\p{-\hvec{d}_\text{rt} } G_\text{t}\p{\hvec{d}_\text{rt} } }, \label{eq:a_d} \\
    a_\text{a} =&d_\text{rs}^{-1} d_\text{st}^{-1} \sqrt{ G_\text{r}\p{-\hvec{d}_\text{rs} } G_\text{t}\p{\hvec{d}_\text{st} } \frac{\bar{\sigma} }{4\pi}} \label{eq:a_a}
\end{align}
with $G_y = \frac{2 \pi \eta \vec{l}\H_y \vec{l}_y}{\lambda^2}$ is the gain of the $y$'th antenna, $\hvec{p}_\text{r}\p{\hvec{d}_\text{rs}, \hvec{d}_\text{st}} =  \mat{S}\p{\hvec{d}_\text{rs}, \hvec{d}_\text{st}} \hvec{l}_\text{t}\p{\hvec{d}_\text{st}} \sqrt{\frac{4\pi}{\sigma}}$, and 
\begin{equation}
    \bar{\sigma} = 4\pi \norm{\mat{S}\p{\hvec{d}_\text{rs}, \hvec{d}_\text{st}} \hvec{l}_\text{t}\p{\hvec{d}_\text{st}}}_2^2 \label{eq:RCS}
\end{equation}
is the the \gls{rcs} of the \gls{ris}.
%$a_\text{d} =  d_\text{rt}^{-1}\sqrt{G_\text{r}\p{-\hvec{d}_\text{rt} } G_\text{t}\p{\hvec{d}_\text{rt} } }$, $a_\text{a} =d_\text{rs}^{-1} d_\text{st}^{-1} \sqrt{ G_\text{r}\p{-\hvec{d}_\text{rs} } G_\text{t}\p{\hvec{d}_\text{st} } \frac{\sigma }{4\pi}}$, $G_y = \frac{2 \pi \eta \vec{l}\H_y \vec{l}_y}{\lambda^2}$ is the gain of the $y$'th antenna, $\sigma = 4\pi \norm{\mat{S}\p{\hvec{d}_\text{rs}, \hvec{d}_\text{st}} \hvec{l}_\text{t}\p{\hvec{d}_\text{st}}}_2^2$ is the \gls{rcs} of the \gls{ris}, and $\hvec{p}_\text{r}\p{\hvec{d}_\text{rs}, \hvec{d}_\text{st}} =  \mat{S}\p{\hvec{d}_\text{rs}, \hvec{d}_\text{st}} \hvec{l}_\text{t}\p{\hvec{d}_\text{st}} \sqrt{\frac{4\pi}{\sigma}}$. 

With the goal of analyzing the wave manipulating capabilities of the \gls{ris}, we define the effective scattering, $A_\text{c}$, and the effective \gls{rcs}, $\sigma$, of the \gls{ris} as
    \begin{align}
    A_\text{c} &= \hvec{p}\H_\text{out} \mat{S}\p{\hvec{d}_\text{rs}, \hvec{d}_\text{st}} \hvec{p}_\text{in}, \label{eq:opt1Variable}\\
    \sigma &= \abs{A_\text{c}}^2 4\pi, \label{eq:effRCS}
    \end{align}
where $\hvec{p}_\text{in}$ and $\hvec{p}_\text{out}$ are the polarization vectors of in- and out-going planewaves. Observe that setting 
\begin{align}
\hvec{p}_\text{out} &=  \e{-\iu k \p{d_\text{rt}- d_\text{st}} }  \e{\iu \angle \p{ \hvec{l}\H_\text{r}\p{-\hvec{d}_\text{rt}} \hvec{l}_\text{t}\p{\hvec{d}_\text{rt}}}} \hvec{l}_\text{r}\p{-\hvec{d}_\text{rs}}, \\
\hvec{p}_\text{in} &=  \e{-\iu k d_\text{rs} }  \hvec{l}_\text{t}\p{\hvec{d}_\text{st}}
\end{align}
and maximizing the utility of the \gls{ris} defined as 
\begin{align}
    A = \Re{A_\text{c}} = \Re{\hvec{p}\H_\text{out} \mat{S}\p{\hvec{d}_\text{rs}, \hvec{d}_\text{st}} \hvec{p}_\text{in}} \label{eq:optVariable}
\end{align}
maximizes the degree to which the signal through the \gls{ris}-assisted path is phase-aligned with the signal through the direct path. As such, the utility $A$ is a metric of degree to which the \gls{ris} is capable of producing polarization- and phase-rotations. In the specific case of independence between scattering amplitude and phase, maximization of the utility $A$ is equivalent of maximizing the signal power. The utility $A$ and \eqref{eq:optVariable} are used as the main optimization objective for the single-element analysis.
\iffalse
Assuming  \eqref{eq:optimizationObjectiveF} is maximized by finding a configuration which maximizes $\sigma$ and adjusts the phase such that $ \hvec{l}\H_\text{r}\p{-\hvec{d}_\text{rt}} \hvec{l}_\text{t}\p{\hvec{d}_\text{rt}} \e{-\iu k d_\text{rt}}  = \hvec{l}\H_\text{r}\p{-\hvec{d}_\text{rs}}  \hvec{p}_\text{r} \p{\hvec{d}_\text{rs}, \hvec{d}_\text{st}} \e{-\iu k \p{d_\text{rs} + d_\text{st}}}$. This is equivalent to maximizing
\begin{align}
A = \Re{\hvec{p}\H_\text{out} \mat{S}\p{\hvec{d}_\text{rs}, \hvec{d}_\text{st}} \hvec{p}_\text{in}} 
\end{align}
where \pre{I've been thinking about this objective function... because in the introduction we talk about how phase and amplitude are not independent, but then formulate an objective function that is optimal only when that condition is met}
\begin{align}
\hvec{p}_\text{out} &=  \e{-\iu k \p{d_\text{rt}- d_\text{st}} }  \e{\iu \angle \p{ \hvec{l}\H_\text{r}\p{-\hvec{d}_\text{rt}} \hvec{l}_\text{t}\p{\hvec{d}_\text{rt}}}} \hvec{l}_\text{r}\p{-\hvec{d}_\text{rs}}, \\
\hvec{p}_\text{in} &=  \e{-\iu k d_\text{rs} }  \hvec{l}_\text{t}\p{\hvec{d}_\text{st}}
\end{align}
are the polarization vectors of the ingoing and outgoing waves. Although maximization of \eqref{eq:optVariable} only yields the optimal configuration under the assumption of independence between scattering amplitude and phase, it serves as a good metric for the utility of the \gls{ris} and is as such used as the main optimization objective for the single-element analysis.
\fi

The remaining part of the modeling is thus to derive an expression for the scattering matrix $ \mat{S}\p{\hvec{r}_\text{out}, \hvec{r}_\text{in}}$.

%---------------------------------------------------------------
% RIS MODEL
%---------------------------------------------------------------
\subsection{RIS model and scattered field}

The electromagnetic field scattered by the \gls{ris} can be computed by inserted its current distribution into \eqref{eq:greensFormulation}---or into \eqref{eq:farfieldPlanewave} for far field radiation. However, the exact current distribution is a complicated function of the physical structure of the particles, being hence impractical. As the particles are electronically small, though, their scattering response is predominantly dipolar, and therefore we can replace them by a set of electric and effective magnetic dipole moments. This approximation allows us to \textit{i)} assume the Green's matrices are approximately constant over the volume of a single particle, and \textit{ii)} calculate the scattered field as the sum of the contribution of each effective dipole. Then, \eqref{eq:greensFormulation} yields
\begin{align}
\begin{bmatrix}\vec{e}_\text{s}(\vec{x}+\veci{x})  \\ \vec{h}_\text{s}(\vec{x}+\veci{x})  \end{bmatrix} &\approx \sum_{n=1}^N \begin{bmatrix}\vec{e}_n(\vec{x}+\veci{x})  \\ \vec{h}_n(\vec{x}+\veci{x})  \end{bmatrix}  \notag \\
 &= \sum_{n=1}^N \mat{G}\p{\vec{x} + \veci{x} - \vec{x}_n}  \begin{bmatrix}  \vec{j}_n   \\
  \vec{m}_n\end{bmatrix}, \label{eq:greensMatrix4}
\end{align}
where $\vec{e}_\text{s}$ and $\vec{h}_\text{s}$ are the fields scattered by the whole \gls{ris}, $\vec{e}_n$ and $\vec{h}_n$ are the fields scattered by the $n$-th particle, with $N$ the total number of particle composing the \gls{ris}, and $\vec{j}_n$ and $\vec{m}_n$ are the currents carried by the $n$-th particle. Note that, with the dipolar approximation, $\vec{j}_n$ and $\vec{m}_n$ can be seen as the discretization of the current distributions $\veci{j}$ and $\veci{m}$ at the points $\vec{x}_n$ (particles positions). It is important to also note that, as $\mat{G}(\cdot)$ is inversely proportional to the distance (c.f. \eqref{eq:Gee}-\eqref{eq:Gem}), the approximation of $\mat{G}(\cdot)$ being constant over the volume of the particle is only valid if that is significantly smaller than the inter-particle spacing when computing the interacting field between them.  

Equation \eqref{eq:greensMatrix4} provides the field scattered by the \gls{ris} in terms of the currents, so the next step is deriving the latter. When the \gls{ris} particles interact with an electromagnetic field, a current density is induced in them. These induced currents are commonly modelled as proportional to the \textit{acting field} \cite{jackson_classical_1999,achouri_electromagnetic_2021,collin_field_1991,yatsenko_electromagnetic_2000}, understanding by acting field not only the one impinging from the transmitter but also the fields generated by adjacent particles. Therefore, the field acting upon the $i$'th particle---$\vec{e}_{\text{a}i}$ and $\vec{h}_{\text{a}i}$---is as such given by
\begin{align}
\begin{bmatrix}
\vec{e}_{\text{a}i}(\vec{x}_i) \\ \vec{h}_{\text{a}i}(\vec{x}_i)
\end{bmatrix} = \begin{bmatrix}
\vec{e}_\text{e}(\vec{x}_i) \\ \vec{h}_\text{e}(\vec{x}_i)
\end{bmatrix} + \sum_{n\neq i} \begin{bmatrix}
\vec{e}_n(\vec{x}_i) \\ \vec{h}_n(\vec{x}_i)
\end{bmatrix}, \label{eq:ActingField}
\end{align}
where $\vec{e}_\text{e},\vec{h}_\text{e}$ is the external field, i.e., that coming from the transmitter or any reflection in the environment. In line with \eqref{eq:farfieldPlanewave}, the external field is in general given as a distribution of plane waves under farfield conditions. With impinging planewave distribution $\veci{e}_\text{in}(\hvec{r}_\text{in})$, given as in \eqref{eq:eOut}, the external field is given as
\begin{align}
    \begin{bmatrix}
        \vec{e}_\text{e}\p{\vec{x}_i} \\
        \vec{h}_\text{e}\p{\vec{x}_i}
    \end{bmatrix} &= \e{-\iu k  \hvec{r}_\text{in}\T\vec{x}_i } \begin{bmatrix}
    \eye_3 & \mat{0}_{3\times 3} \\
     \mat{0}_{3\times 3} & \eye_3 \eta^{-1} 
    \end{bmatrix}  \mat{o} \veci{e}_\text{in}(\hvec{r}_\text{in}) \label{eq:ExternalField}\\
    \mat{O} &= \begin{bmatrix}
\mat{I}_3 & -\mat{P}_2(\hvec{r}_\text{in})
\end{bmatrix}\T.
\end{align}
% $\veci{e}_\text{in}(\hvec{r}_\text{in})$ as in \eqref{eq:eOut} and
%\begin{align}
%    \mat{O} = \begin{bmatrix}
%1 & 0 & 0 &  0 & \p{\hvec{r}_\text{in}}_3  & -\p{\hvec{x}_\text{in}}_2 \\
%0 & 1 & 0 & -\p{\hvec{r}_\text{in}}_3 & 0 & \p{\hvec{x}_\text{in}}_1 \\
%0 & 0 & 1 & \p{\hvec{r}_\text{in}}_2 & -\p{\hvec{x}_\text{in}}_1 & 0
%\end{bmatrix}\T.
%\end{align}
%\begin{align}
%    \mat{O} = \begin{bmatrix}
%\mat{I}_3 & -\mat{P}_2(\hvec{r}_\text{in})
%\end{bmatrix}\T.
%\end{align}

With the external field characterized, let define a proportionality factor matrix $\mat{X}\in\mathbb{C}^{6\times 6}$ such that the induced currents in the $i$-th particle are given by
\begin{align}
\begin{bmatrix} \vec{j}_i \\ \vec{m}_i \end{bmatrix} = \mat{X}_i \begin{bmatrix}
\vec{e}_{\text{a}i}(\vec{x}_i) \\ \vec{h}_{\text{a}i}(\vec{x}_i)
\end{bmatrix}. \label{eq:greensMatrix5}
\end{align}
It is important to note that the factor $\mat{X}$ is equivalent to the widely applied polarizability concept apart from a factor $\iu \omega$ \cite{jackson_classical_1999,achouri_electromagnetic_2021,collin_field_1991,yatsenko_electromagnetic_2000}. Thus, due to the similarity between them, we refer here to $\mat{X}$ as the polarizability matrix. Recalling the dipolar particle response, $\mat{X}$ characterizes the excitation of each magnetic and electric dipole moment due to the field strength at each direction, and depends on the specific particle implementation and its configuration. Therefore, we assume $\mat{X}$ is, in general, reconfigurable. 

\begin{remark}[Passive RIS]
    \label{remark:Passive}
    Depending on the physical implementation of the \gls{ris}, different constraints apply to $\mat{X}$. An important one is that, to make the particles lossless and passive (being both a common assumption in wireless communications), then the condition
    \begin{align}
        \frac{\mat{X}_i + \mat{X}_i\H}{2}  = -\mat{X}_i\H \mat{G}_0 \mat{X}_i \label{eq:PassiveCondition}
    \end{align}
    must hold (see Appendix \ref{app:1}). Using the results in \cite[Thm. 11]{Crone1981}, this translates into the structural constraint
    \begin{align}
        \mat{X}_i = \p{-\mat{G}_0}^{-\frac{1}{2}} \frac{ \mat{U}_i  + \eye_6}{2}  \p{-\mat{G}_0}^{-\frac{1}{2}}, \label{eq:polarizabilitySolution}
    \end{align}
    where $\mat{U}_i \in \mathbb{C}^{6\times 6}$ is any unitary matrix.
\end{remark}

Considering that all the particles meet \eqref{eq:polarizabilitySolution}, the currents induced at the $N$ particles can be obtained from \eqref{eq:ActingField}-\eqref{eq:greensMatrix5}, yielding a system of $6N$ equations and unknowns. Solving it and introducing the solution into \eqref{eq:greensMatrix4}, the planewave spectrum scattered by the \gls{ris} is obtained as
\begin{align}
\vec{e}_\text{s}\p{\hvec{r}_\text{out}} =&\frac{ \iu 3 }{4 k } \mat{P}\mat{H} 
\p{\eye_{6N} -  \frac{ \iu 3\pi}{k} \p{\mat{V} + \eye_{6N}} \mat{M}}^{-1}  \notag \\
&\times \p{\mat{V} + \eye_{6N}}  \mat{T} \mat{O} \vec{e}_\text{in}(\hvec{r}_\text{in}),\label{eq:scatteredWave}
\end{align} %\frac{ \iu 3 \e{-\iu k d_\text{rs}}}{4 k d_\text{rs} }
where the different involved matrices are defined as
\begin{align}
&\mat{H} =  \begin{bmatrix}
\e{\iu k \hvec{r}_\text{out}\T\vec{x}_1} \eye_{6} &  \e{\iu k \hvec{r}_\text{out}\T\vec{x}_2} \eye_{6} & \dots & \e{\iu k \hvec{r}_\text{out}\T\vec{x}_N} \eye_{6}\end{bmatrix}, \notag \\
&\mat{T} = \begin{bmatrix}  \e{\iu k \hvec{r}_\text{in}\T \vec{x}_1  }\eye_6 & \e{\iu k \hvec{r}_\text{in}\T \vec{x}_2  }\eye_6 & \cdots & \e{\iu k \hvec{r}_\text{in}\T \vec{x}_N  }\eye_6\end{bmatrix}\H, \notag \\
&\mat{V} = \begin{bmatrix} 
\mat{U}_1 	& 0 		& \cdots 	& 0 \\
0 		& \mat{U}_2 	& \cdots 	& 0 \\
\vdots 		& \vdots 	& \ddots 	& \vdots \\
0 		& 0 		& \cdots 	& \mat{U}_N  
\end{bmatrix}, \notag \\
&\mat{M} = \begin{bmatrix}
0 				& \mat{G}'\p{\vec{x}_1-\vec{x}_2}	& \cdots 	& \mat{G}'\p{\vec{x}_1-\vec{x}_N} \\
\mat{G}'\p{\vec{x}_2-\vec{x}_1} 	& 0 					& \cdots 	& \mat{G}'\p{\vec{x}_2-\vec{x}_N} \\
\vdots 				& \vdots 				& \ddots 	& \vdots \\
\mat{G}'\p{\vec{x}_N-\vec{x}_1} 	& \mat{G}'\p{\vec{x}_N-\vec{x}_2} 	& \cdots 	& 0 
\end{bmatrix}, \notag \\
&\mat{G}'\p{\vec{x}} =\begin{bmatrix}
        \mat{G}_\text{ee}\p{\vec{x}}& \mat{G}_\text{em}\p{\vec{x}} \\
        \mat{G}_\text{me}\p{\vec{x}}& \mat{G}_\text{mm}\p{\vec{x}}
    \end{bmatrix}. \notag 
\end{align}

Comparing \eqref{eq:eOut} and \eqref{eq:scatteredWave} directly yields that the scattering matrix of the \gls{ris} is given as
\begin{align}
\mat{S}(\hvec{r}_\text{out},\hvec{r}_\text{in}) =& \frac{\iu 3}{ 4 k}\mat{P}\mat{H} 
\p{\eye_{6N} -  \frac{ \iu 3\pi}{k} \p{\mat{V} + \eye_{6N}} \mat{M}}^{-1} \notag \\
&\times\p{\mat{V} + \eye_{6N}}  \mat{T} \mat{O}, \label{eq:S_normalized}
\end{align}
completing the end-to-end channel modelling. 

\begin{remark}[Reciprocal RIS]
\label{remark:Reciprocity}
    For many communication systems, and especially time-division-duplex multiple-input-multiple-output systems, channel reciprocity is exploited to significantly reduce system overhead. For \gls{ris} systems, channel reciprocity is not a given. Defining the submatrices of $\mat{X}_i$ as
\begin{align}
    \mat{X}_i = \begin{bmatrix} \mat{X}_\textnormal{ee} & \mat{X}_\textnormal{em} \\
    \mat{X}_\textnormal{me} & \mat{X}_\textnormal{mm}\end{bmatrix},\label{eq:Xsubmatrices}
\end{align}
then channel reciprocity holds if $\mat{X}_\textnormal{ee} = \mat{X}_\textnormal{ee}\T$, $\mat{X}_\textnormal{mm} = \mat{X}_\textnormal{mm}\T$, $\mat{X}_\textnormal{em} = -\mat{X}_\textnormal{me}\T$ (see Appendix \ref{app:1}), which is always satisfied if $\mat{X}_i$ is a diagonal matrix. Due to \eqref{eq:polarizabilitySolution}, identical symmetry constraints carries over to the four three-by-three submatrices of $\mat{U}_i$.
\end{remark}

\section{Amplitude-phase relationship for passive and lossless particles}\label{sec:properties}
\begin{figure}[t]
    \centering
    \includegraphics[width=0.9\linewidth,trim = {1cm 0 1cm 0}]{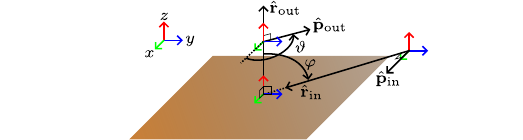}
    \caption{Illustration of the geometry used in the single-element analysis.}
    \label{fig:geometry_single_element}
\end{figure}

\subsection{RIS utility in terms of phase-shift, polarization and direction}

The \gls{ris} scattering matrix $\mat{S}$ can be reconfigured by modifying the particles polarizabilities in order to maximize the radiation towards a given direction or to change the polarization. To get insight in the relationship between arrival/departure directions, amplitude, phase and polarization of the outgoing wave, we first particularize \eqref{eq:S_normalized} for a single particle\footnote{This implies that $\mat{M}=\vec{0}, \mat{H}=\mat{T}=\mat{I}_6$.}, yielding
\begin{align}
\mat{s}(\hvec{r}_\text{out},\hvec{r}_\text{in}) &= \frac{ \iu 3 }{4 k }
\mat{P} \p{\mat{U} + \eye_6} \mat{o} \label{eq:singleElementScattering}
\end{align}
Introducing this into the utility function \eqref{eq:optVariable} leads to
\begin{align}
    A &= \frac{ 3 }{4 k} \Re{\iu \hvec{p}_\text{out}\H \mat{P} \p{\mat{U} + \eye_6} \mat{o} \hvec{p}_\text{in} } \notag \\
     &= \frac{ 3}{4 k} \Re{ \Tr{  \mat{U}  \mat{U}' \mat{\Sigma} \mat{V}\H } + \iu \hvec{p}_\text{out}\H  \mat{P} \mat{o} \hvec{p}_\text{in} } \notag
\end{align}
where $ \mat{U}' \mat{\Sigma} \mat{V}\H = \iu \mat{o} \hvec{p}_\text{in}\hvec{p}_\text{out}\H \mat{P} $ is the singular value decomposition. Trivially, the utility $A = \Re{A_\text{c}}$ is maximized when $\mat{U} = \mat{V} {\mat{U}'}\H$. The effective scattering, \eqref{eq:opt1Variable}, under maximum utility is given as
\begin{align}
    A_\text{c}^*  \overset{(a)}{=} \frac{3}{2 k}  + \frac{3 \iu \hvec{p}_\text{out}\H  \mat{P} \mat{o} \hvec{p}_\text{in}}{4 k}   \implies A^* = \Re{A_\text{c}^*}  \leq  \frac{3}{k}, \label{eq:maxUtility}
\end{align}
where $(a)$ uses the fact that, due to the structure of the problem, $\p{\mat{\Sigma}}_{n,n}=2$ for $n= 1$ and $0$ otherwise. Note also that the solution $\mat{U} = \mat{V} {\mat{U}'}\H$ is not unique, and that it does not necessarily meet the conditions in Remark \ref{remark:Reciprocity} (and hence the system may not be reciprocal). 

%\begin{figure*}[t]
%    \centering
%    \includegraphics[width=\linewidth]{figures/max_reflection.pdf}
%    \caption{Heatmaps showing the maximum utility $A^*$ provided by a single element, ignoring the reciprocity constraints. \textit{Left}: The utility under parallel polarization as a function of phase-shift $\rho$ and angular separation $\varphi$. \textit{Right}: The utility as a function of the polarization rotation of the reflected wave $\vartheta$ and the phase-shift $\rho$ for a constant separation angle of $\varphi = \frac{\pi}{2}$.}
%    \label{fig:single_element_utility}
%\end{figure*}

As assumed in most of the works on \gls{ris}-assisted communications, one would expect maximum utility (i.e, $A=3/k$) regardless the arrival/departure directions and the desired phase shift. However, we shall see that this is not met in practice. To further analyze this, we consider the case where the particle is excited by one plane wave with single polarization. Specifically, we assume direction of arrival and departure given by
\begin{align}
\setlength\arraycolsep{3pt}
\hvec{r}_\text{in} = \begin{bmatrix}
0 & -\es{\varphi} & -\ec{\varphi}
\end{bmatrix}\T,\quad \hvec{r}_\text{out} = \begin{bmatrix}
0 & 0 & 1
\end{bmatrix}\T,\label{eq:Direction}
\end{align}
and polarization vectors
\begin{align}
\setlength\arraycolsep{3pt}
\hvec{p}_\text{in} = \begin{bmatrix}
1 & 0 & 0
\end{bmatrix}\T, \quad
\hvec{p}_\text{out} = \e{\iu \rho}\begin{bmatrix}
\ec{\vartheta} & \es{\vartheta} & 0
\end{bmatrix}\T, \label{eq:polarization}
\end{align}
where $\rho\in[0,2\pi)$ is the phase shift between the ingoing and outgoing waves. See Fig. \ref{fig:geometry_single_element} for a graphical illustration of the scenario. 

In this scenario the effective scattering under maximum utility is given as
\begin{align}
    A^*_\text{c} =  \frac{3}{2 k} + \frac{3}{4 k}\iu\e{-\iu \rho} \ec{\vartheta}\p{\ec{\varphi} - 1}. \label{eq:maxScattering}
\end{align}
It is clearly observed that the scattering amplitude is a function of the spatial angles, the polarization rotation and the phase shift, invalidating the common assumption of independent amplitude and phase-shift. In fact, even for the case of parallel polarization, $\vartheta = 0$, the utility is maximized only for a relatively small subset of phase shifts and angular separations. Of special relevance are the two specific cases: \textit{i)} orthogonal polarization---$\vartheta = \frac{\pi}{2}$, $\varphi\in [0, 2\pi), \rho \in [0, 2\pi)$---and \textit{ii)} same angle---$\varphi = 0$, $\vartheta\in [0, 2\pi), \rho \in [0, 2\pi)$--- where the maximum utility is independent of the phase-shift.

\subsection{Constant amplitude phase-shifting under reciprocity constraint}
\label{subsection:Constant_phase_shift}

Following with the previous case in \eqref{eq:Direction}-\eqref{eq:polarization}, we aim to analyze now under which conditions it is correct to assume independence between amplitude and phase-shift. To that end, in the following we consider the effective scattering $A_\text{c}$ particularized for parallel and orthogonal polarizations:
\subsubsection{Parallel polarization, $\vartheta=0$} In this case, from \eqref{eq:singleElementScattering} and \eqref{eq:Direction}-\eqref{eq:polarization} we have
\begin{align}
\Eval{A_\text{c}}{\vartheta = 0}{} 
=  \frac{ 3  }{4 k}\e{-\iu \p{\rho+\frac{\pi}{2}} } \begin{bmatrix}
-u_{1, 5} - u_{5, 5} - 1 \\
u_{1, 6} + u_{5, 6} \\
u_{1, 1} + u_{5, 1} + 1
\end{bmatrix}\T \begin{bmatrix}
\ec{\varphi} \\
\es{\varphi} \\
1
\end{bmatrix},\label{eq:linearPolarizedScattering}
\end{align}
where $u_{m,n}=(\mat{U})_{m,n}$. Consider first the illustrative example of same direction of arrival/departure ($\varphi = 0$). In this scenario, we have
\begin{align}
\Eval{A_\text{c}}{\varphi = 0, \vartheta = 0}{} &=\frac{3 \e{-\iu \p{\rho+\frac{\pi}{2}}}}{4 k} \p{u_{1,1}-u_{1,5}+u_{5,1}-u_{5,5}},
\end{align}
which, under the reciprocity constraints in Remark \ref{remark:Reciprocity}, attains its maximum value of $\frac{3}{2 k}$ when either $u_{1,1} = -u_{5,5} = \e{\iu \p{\rho +\frac{\pi}{2}}}$ and $u_{5,1} = u_{1,5} = 0$ (see \eqref{eq:unitaryExampleBroadsideParallel} in Appendix \ref{app:2}), or $u_{5,1} = -u_{1,5} = \e{\iu \p{\rho +\frac{\pi}{2}}}$ and $u_{1,1} = u_{5,5} = 0$. In these cases $\Eval{A_\text{c}}{\varphi = 0, \vartheta = 0}{}$ is not a function of $\rho$ and thus achieves constant amplitude phase-shifting. For angles $\varphi \neq 0$, the amplitude however varies with $\rho$. As an example, consider the case of orthogonal directions $\varphi = \pi/2$, which leads to %varies and the phase of the reflected wave does not equal $\rho$. 
\begin{align}
\Eval{A_\text{c}}{\varphi = \frac{\pi}{2}, \vartheta = 0}{} &= \frac{3 \e{-\iu \p{\rho + \frac{\pi}{2}}}}{4k}  \p{1 + u_{1,1} + u_{1,6} + u_{5,1} + u_{5,6}}
\end{align}
where the real part can be maximized by setting $u_{1,1} = u_{5,6} = \e{\iu\p{\rho + \frac{\pi}{2}}}$ and $u_{1,6} = u_{5,1} = 0$, yielding
\begin{align}
\Eval{A_\text{c}}{\varphi = \frac{\pi}{2}, \vartheta = 0}{} &= \frac{3}{4k}  \p{2 + \e{-\iu \p{\rho + \frac{\pi}{2}}}}
\end{align}
which shows that the utility for a direction orthogonal to the impinging wave can be stronger than utility achieved for a direction parallel to the impinging wave. Maximum scattering is observed when the reflected wave is delayed by a quarter cycle, $\rho = -\frac{\pi}{2}$, and minimum scattering is observed for a phase-shift of $\rho = \frac{\pi}{2}$. 

Interestingly, scattering amplitude can however be sacrificed to create constant amplitude phase-shifting, this is observed for the configuration $\mat{U}$ given in \eqref{eq:unitaryExampleOrthogonalParallel} in Appendix \ref{app:2}, for which $\Eval{A_\text{c}}{\varphi = \frac{\pi}{2}, \vartheta = 0}{} = 3/(4k)$, being independent of $\rho$. 

\begin{remark}
    From \eqref{eq:linearPolarizedScattering} we observe that, to achieve constant amplitude phase-shifting for every direction, the dependence on $\rho$ and $\varphi$ needs to be canceled out. This requires $u_{1,5}= -(u_{5,5}+1)$, $u_{1,6}=-u_{5,6}$ and $u_{1,1}+u_{5,1} = \alpha\e{-\iu (\rho + \pi/2)} - 1$ with $\alpha\in\mathbb{R}$. Since $\mat{U}$ is unitary for passive systems, hence $|u_{1,1}|^2+|u_{5,1}|^2\leq 1$, yielding $\alpha =0$ as the only option, which implies no scattering. Therefore, it is not physically possible to achieve independence between phase and amplitude for parallel polarization.
\end{remark}

\subsubsection{Orthogonal polarization, $\vartheta = \frac{\pi}{2}$} \label{subsubsec:Orthogonal_polarization} In this case, we get
\begin{align}
\Eval{A_\text{c}}{\vartheta = \frac{\pi}{2}}{} &=  \frac{ 3 }{4k}\e{-\iu \p{\rho+\frac{\pi}{2}} } \begin{bmatrix}
 u_{4, 5} - u_{2, 5} \\
 u_{2, 6} - u_{4, 6} \\
 u_{2, 1} - u_{4, 1}
\end{bmatrix}\T \begin{bmatrix}
\ec{\varphi} \\
\es{\varphi} \\
1
\end{bmatrix}
\end{align}
where it can be seen that constant amplitude phase-shifting is achieved for any one angle $\varphi = \phi$ by setting $u_{2, 1} = \e{\iu \p{\rho+\frac{\pi}{2}} }$, $u_{4, 5} = \ec{\phi}\e{\iu \p{\rho+\frac{\pi}{2}} }$, and $u_{4, 6} = -\es{\phi}\e{\iu \p{\rho+\frac{\pi}{2}} }$ in which case $\Eval{A_\text{c}}{\vartheta = \frac{\pi}{2}, \varphi = \phi}{} =3/(2k)$. When an element is tuned for a single angle $\varphi = \phi$ in this manor, off-angle reflections, $\varphi \neq \phi$, does not provide constant amplitude phase-shifting. Finally, one can show that constant amplitude phase-shifting is possible for all on-axis directions simultaneously with the configuration given in \eqref{eq:unitaryExampleOrthogonalPolarization} in Appendix \ref{app:2}, yielding $\Eval{A_\text{c}}{\vartheta = \frac{\pi}{2}}{} = 3 \sqrt{2}/(4 k)$, which is completely independent of $\varphi$.

\section{Impact of phase, amplitude and angle dependency in communications }\label{sec:impact_commun}
%----------------------------------------------------------------------------
% IMPACT ON CHANNEL ESTIMATION AND CASCADED CHANNEL
%----------------------------------------------------------------------------

\subsection{Implications in cascaded channel model and channel estimation}

In the following, we aim to evaluate the impact on wireless communications of the tight dependency between phase, amplitude, polarization and angle highlighted in the previous section.  To that end, we analyze three different scenarios: \textit{i)} a single particle in a line-of-sight channel, \textit{ii)} a single particle in a random uniform planar scattering environment, known as Clarke's two-dimensional model \cite[Sec. 5.4]{parsons_mobile_2000}, and \textit{iii)} a linear array of elements under Clarke's model.

\subsubsection{Single element, line-of-sight}
The \gls{ris}-assisted channel\footnote{The direct link between transmitter and receiver is here ignored.} is given by \eqref{eq:recSignal} and \eqref{eq:LOSChannel} as
\begin{align}
    h = \frac{\eta}{2\lambda}  \frac{\e{-\iu k (d_\text{rs} + d_\text{st})}}{ d_\text{rs} d_\text{st}} \vec{l}\H_\text{r}\p{-\hvec{d}_\text{rs}} \mat{S}\p{\hvec{d}_\text{rs}, \hvec{d}_\text{st}} \vec{l}_\text{t}\p{\hvec{d}_\text{st}}. \label{eq:h_los_commun}
\end{align}
With the transmitter and receiver located in the $yz$-plane at angles $\phi_1$ and $\phi_2$ w.r.t. the $z$-axis respectively, we have
\begin{equation}
\begin{aligned}
\vec{l}_\text{t}\p{\hvec{d}_\text{st}} &= a_\text{t} \setlength\arraycolsep{3pt}\begin{bmatrix}
1 & 0 & 0
\end{bmatrix}\T, & \vec{l}_\text{r}\p{-\hvec{d}_\text{rs}} &= a_\text{r}\setlength\arraycolsep{3pt} \begin{bmatrix}
1 & 0 & 0
\end{bmatrix}\T, \\
\hvec{d}_\text{st} &= \begin{bmatrix}
0 \\ -\es{\phi_1} \\ -\ec{\phi_1}
\end{bmatrix}, & \hvec{d}_\text{rs} &= \begin{bmatrix}
0 \\ \es{\phi_2} \\ \ec{\phi_2}
\end{bmatrix},
\end{aligned}
\end{equation}
with $a_\text{t}$ and $a_\text{r}$ as in \eqref{eq:a_d} and \eqref{eq:a_a}. Consider also a parallel polarization as in \eqref{eq:polarization} with $\vartheta = 0$. Introducing these into \eqref{eq:h_los_commun} and considering the scattering matrix \eqref{eq:singleElementScattering}, the resulting channel for a configuration $\mat{U}$ for parallel directions \eqref{eq:unitaryExampleBroadsideParallel}, denoted by $h_\text{p}$, and for orthogonal directions ($\mat{U}$ as in \eqref{eq:unitaryExampleOrthogonalParallel}), denoted by $h_\text{o}$, are given by 
\begin{align}
    h_\text{p}\p{\rho} &= \frac{ 3 \eta }{16 \pi}  \frac{\e{-\iu k (d_\text{rs} + d_\text{st})}}{ d_\text{rs} d_\text{st}}  \p{g_{\text{p}1} \e{\iu \rho}  - \iu g_{\text{p}2} } a_\text{r} a_\text{t}, \\
    h_\text{o}\p{\rho} &= \frac{ 3 \eta }{16 \pi}  \frac{\e{-\iu k (d_\text{rs} + d_\text{st})}}{ d_\text{rs} d_\text{st}}  \p{\e{\iu \rho}  - \iu  g_{\text{o}}  } a_\text{r} a_\text{t},
\end{align}
where $g_{\text{p}1} = 1 + \ec{\phi_1}\ec{\phi_2}$, $g_{\text{p}2} = 1 - \ec{\phi_1}\ec{\phi_2}$ and $g_{\text{o}} = 1 - \p{\es{\phi_1} + \ec{\phi_1}} \p{\es{\phi_2} + \ec{\phi_2}} $. Interestingly, observe that, for two different phase-shifts $\rho_1$ and $\rho_2$, the following holds:
\begin{align}
    h_\text{p}\p{\rho_1} &= h_\text{p}\p{\rho_2} \frac{g_{\text{p}1} \e{\iu \rho_1} - \iu  g_{\text{p}2} }{g_{\text{p}1} \e{\iu \rho_2} - \iu  g_{\text{p}2} }, \label{eq:hp_rel1}\\
    h_\text{o}\p{\rho_1} &= h_\text{o}\p{\rho_2}  \frac{\e{\iu \rho_1} -\iu  g_{\text{o}}  }{\e{\iu \rho_2}  - \iu  g_{\text{o}} } \label{eq:hp_rel2}. 
\end{align}

The above relation has a strong implication in wireless communications. Usually, to estimate the end-to-end channel (\gls{ris}-assisted path), the \gls{ris} is fixed to a certain configuration---i.e., a certain $\rho$, and the cascaded channel is then estimated, assuming this estimation of the transmitter-to-\gls{ris} and \gls{ris}-to-receiver channels remains valid for others phase-shifts. However, looking at \eqref{eq:hp_rel1}-\eqref{eq:hp_rel2}, we see that predicting $h\p{\rho_1}$ based on $h\p{\rho_2}$ requires not only information about $h\p{\rho_2}$ itself, but also knowledge about the angular positions of the transmitter and receiver (equivalently, $g_{\text{p},1}, g_{\text{p},2}$ and $g_{\text{o}}$). In effect, this means that even for a single element \gls{ris}, multiple pilot symbols are necessary to fully know the channel due to the lack of constant amplitude phase-shifting for all angles simultaneously.

Assume now an orthogonal polarization ($\varrho = \pi/2$ in \eqref{eq:polarization}), and a configuration matrix $\mat{U}$ as in \eqref{eq:unitaryExampleOrthogonalPolarization}. Setting $ \vec{l}_\text{r}\p{-\hvec{d}_\text{rs}} = a_\text{r}\begin{bmatrix}
0 & 1 & 0
\end{bmatrix}\T$ yields a received signal
\begin{align}
    h_\text{r}\p{\rho} &= \frac{3 \eta \sqrt{2}}{32 \pi} \ec{\phi_2} \p{1 + \ec{\phi_2}} \e{\iu \rho} a_\text{r} a_\text{t},
\end{align}
where there is no dependence on the angular direction due to the constant amplitude phase-shifting achieved for orthogonal polarizations (c.f. Section \ref{subsubsec:Orthogonal_polarization}). Then, for two different phase-shifts $\rho_1$ and $\rho_2$, we get now
\begin{align}
    h_\text{r}\p{\rho_1} &= v_\text{r}\p{\rho_2} \e{\iu \p{\rho_1 - \rho_2}},
\end{align}
which implies that one only needs knowledge of $v_\text{r}\p{\rho_2}$ to fully know the channel for any phase-shift $\rho$. In contrary to the parallel polarization case, full channel information across different reflection phases can be achieved by sending only a single known pilot symbol.

\subsubsection{Single element, multiple paths}

To analyze this case, we follow the same steps us before, but assuming now that the transmitter-to-\gls{ris} channel is no longer line-of-sight but describer according to Clarke's model. Hence, the wave impinging upon the \gls{ris} is a sum of $M$, $x$-polarized plane waves, all travelling in the $yz$-plane, i.e., 
\begin{equation}
    \mat{H}_\text{st}\p{\hvec{r}, \hvec{r}_\text{t}} \vec{l}_\text{t}\p{\hvec{r}_\text{t}} = \sum_{n=0}^{M} \begin{bmatrix}
\alpha_n & 0 & 0
\end{bmatrix}\T  \delta\p{\hvec{r}_\text{t} - \hvec{r}_n}, 
\end{equation}
where $\alpha_n\in\mathbb{C}$ is a set of circularly symmetric i.i.d. random variables, and $\hvec{r}_n =\setlength\arraycolsep{2pt} \begin{bmatrix}
0 & -\es{\phi_n} & -\ec{\phi_n}
\end{bmatrix}\T$ with $\phi_n \sim \mathcal{U}\p{0, 2\pi}$. From \eqref{eq:ha_vector} and \eqref{eq:recSignal},
the \gls{ris}-assisted channel is thus given by
\begin{align}
    h = \frac{\eta }{2\lambda } \frac{ \e{-\iu k d_\text{rs}}}{ d_\text{rs}} \vec{l}\H_\text{r}\p{-\hvec{d}_\text{rs}} \sum_{n=0}^{M} \mat{S}\p{\hvec{d}_\text{rs}, \hvec{r}_n} \begin{bmatrix}
\alpha_n \\ 0 \\ 0
\end{bmatrix}  \delta\p{\hvec{r}_\text{t} - \hvec{r}_n} .
\end{align}
Particularizing for  $\hvec{d}_\text{rs} =\setlength\arraycolsep{2pt} \begin{bmatrix}
0 & 0 & 1
\end{bmatrix}\T$ and considering parallel polarization, the end-to-end channel the for parallel and orthogonal configurations in \eqref{eq:unitaryExampleBroadsideParallel} and \eqref{eq:unitaryExampleOrthogonalParallel}, respectively, yields
\begin{align}
    h_\text{p}\p{\rho} &= \frac{ 3 \eta }{16 \pi}  \frac{\e{-\iu k d_\text{rs}}}{ d_\text{rs}}   a_\text{r}  \sum_{n=0}^{M} \p{ g_{\text{p}1,n} \e{\iu \rho} -\iu g_{\text{p}2,n}  } \alpha_n ,\label{eq:reflectResponse}\\
    h_\text{o}\p{\rho} &= \frac{ 3 \eta }{16 \pi}  \frac{\e{-\iu k d_\text{rs}}}{ d_\text{rs}}  a_\text{r} \sum_{n=0}^{M}  \p{\e{\iu \rho} -\iu  g_{\text{o},n}  } \alpha_n,
\end{align}
with $g_{\text{p}1,n} = 1 + \ec{\phi_n}$, $g_{\text{p}2,n} = 1 - \ec{\phi_n} $ and $g_{\text{o},n} = 1 - \p{\es{\phi_n} + \ec{\phi_n}}$. 

As $M\rightarrow \infty$, as stated in Clarke's model, the central limit theorem holds and $h_m \sim \mathcal{CN}\p{0, \Sigma_m\p{\rho}^2}$ with $m=\{\text{p},\text{o}\}$ and 
\begin{align}
\Sigma&^2_\text{p}\p{\rho_1, \rho_2} =  \Aver{   h_\text{p}\p{\rho_1}  h_\text{p}\conj\p{\rho_2}   }  \\
&= \p{\frac{3 \eta \abs{a_\text{r}} \Sigma_a}{16 \pi d_\text{rs}}}^2 \frac{ \iu \p{\e{\iu \rho_1}-\e{-\iu \rho_2 }} + 3 \p{1 + \e{\iu \p{\rho_1 - \rho_2 }}}  }{2},  \notag \\
\Sigma&^2_\text{o}\p{\rho_1, \rho_2} =  \Aver{   h_\text{o}\p{\rho_1}  h_\text{o}\conj\p{\rho_2}   }  \\
&= \p{\frac{3 \eta \abs{a_\text{r}} \Sigma_a}{16 \pi d_\text{rs}}}^2 \p{ \iu \p{\e{\iu \rho_1 }+\e{-\iu \rho_2}} +  \e{\iu \p{\rho_1 - \rho_2 }} + 2 }, \notag
\end{align}
where $\Sigma_a^2 = N \Aver{\abs{\alpha_n}^2}$. In effect, the signal is a joint Gaussian distribution across different phase-shifts, given that polarization is preserved. This in turn means that when considering the cascaded channel \eqref{eq:cascadedChannel} as
\begin{align}
h = h_\text{rs} \varphi h_\text{st}\quad \quad \abs{\varphi} \leq 1,
\end{align}
one cannot assume that $h_\text{st}$ and $h_\text{rs}$ remains constant across different $\varphi$. In fact, as is shown here, $h_\text{st}$ and $h_\text{rs}$ are correlated random variables across different phase-shifts, $\varphi$. As anticipated, the randomness of the channel across different phase-shifts has big consequences for signal space channel estimation procedures. This will be further explored next.

\subsubsection{Multiple elements, multiple paths}

Let us generalize the previous scenario to a complete \gls{ris} made from $N$ particles, and consider for analytical simplicity in this example that the elements are sufficiently spaced so that spatial correlation and inter-element coupling is negligible. Then, the received signal in the $m$-th symbol period is the contribution of the $N$ independent elements as
\begin{align}
q_m = \sum_{n=1}^N h_{n}\p{\rho_{n,m}}
\end{align}
where $\rho_{n,m}$ is the phase-shift applied by the $n$-th \gls{ris} element in the $m$-th symbol period.

Assuming we have $M$ periods, we can write the theoretical received signal vector $\tilde{\vec{q}}$ predicted by the cascaded channel model as
\begin{align}
\tilde{\vec{q}} = \mati{\mathcal{P}} \vec{h}, \label{eq:wrongTraining}
\end{align}
where $\p{\mati{\mathcal{P}}}_{n,m} = \e{\iu \rho_{n,m}}$ is the phase-shifting matrix of the \gls{ris} and, as the cascaded model assumes that the ingoing and outgoing channels are independent, $\p{\vec{h}}_n = h_n\p{0}$. Note that we assume the channel remains constant for $m=1,\dots,M$. From \eqref{eq:wrongTraining}, one can draw the conclusion that given $N=M$ and $\mati{\mathcal{P}}$ being full rank, the channel $\vec{h}$ can be estimated as
\begin{align}
\tilde{\vec{h}} =  \mati{\mathcal{P}}^{-1} \tilde{\vec{q}}.
\end{align}
The optimal set of phase-shifts $\rho_n^*$ is thus the set of phases which cancels the phase of $\p{\tilde{\vec{h}}}_n$ such that the product $\p{\tilde{\vec{h}}}_n \e{\iu \rho_n^*} = \abs{\p{\tilde{\vec{h}}}_n}$. We can define therefore the metric $\gamma$ as the power ratio of the actual signal power versus the expected signal power according to the cascaded channel, i.e.,
\begin{align}
\gamma= \p{\frac{  \abs{ \sum_{n=1}^N h_{n}\p{\rho_n^*}} }{ \sum_{n=1}^N\abs{\p{\tilde{\vec{h}}}_n}  } }^2. \label{eq:gamma}
\end{align}

\begin{figure}[t]
    \centering
    \includegraphics{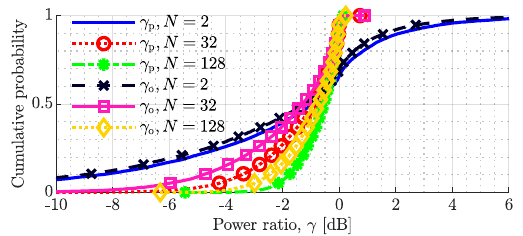}
    \caption{An empirical cumulative distribution function based on Monte Carlo simulations of the power ratio $\gamma$ \eqref{eq:gamma} for the two element configurations; Parallel reflection $\gamma_\text{p}$, and Orthogonal reflection $\gamma_\text{o}$.}
    \label{fig:gamma_cdf}
\end{figure}

This metric is evaluated numerically, and its empirical cumulative distribution function is plotted in Fig. \ref{fig:gamma_cdf} for different \gls{ris} sizes, where $\mati{\mathcal{P}}$ is the discrete Fourier transform matrix, and two configurations---parallel \eqref{eq:unitaryExampleBroadsideParallel} and orthogonal \eqref{eq:unitaryExampleOrthogonalParallel} reflections---are employed \eqref{eq:unitaryExampleOrthogonalParallel}. It is interesting that, as $N$ grows, the variance in both configurations gets smaller while $\Aver{\gamma} \to 1$. Hence, for a sufficiently large array, the estimated signal power tends to the actual signal power.

\begin{figure}[t]
    \centering
    \includegraphics{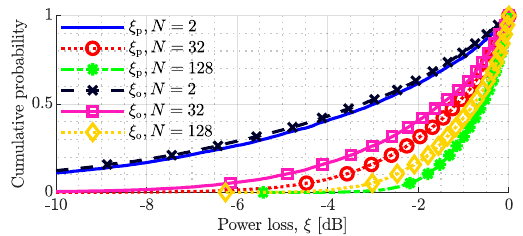}
    \caption{An empirical cumulative distribution function based on Monte Carlo simulations of the power loss $\xi$ \eqref{eq:xi} for the two element configurations; Parallel reflection $\xi_\text{p}$, and Orthogonal reflection $\xi_\text{o}$.}
    \label{fig:xi_cdf}
\end{figure}

Another metric of interest is the power loss $\xi$, defined as
\begin{align}
\xi= \p{\frac{  \abs{ \sum_{n=1}^N h_{n}\p{\rho_n^*}} }{ \abs{\sum_{n=1}^N\p{\vec{h}^*}_n}  } }^2 \label{eq:xi}
\end{align}
where $\vec{h}^*$ is the signal vector given an optimal configuration which maximizes the received signal power and $\rho_n^*$ is again chosen assuming the validity of cascaded channel. Thus, $\xi$ measures how much power is lost when configuring the \gls{ris} based on the cascaded channel estimation. This metric is evaluated in Fig. \ref{fig:xi_cdf}, which shows its empirical cumulative distribution functions, computed using the same $\mati{\mathcal{P}}$ and element configurations as before. The channel $\vec{h}^*$ is obtained by doing a grid search over $\nu = [-\pi; \pi]$ where the phaseshift of the $n$'th element is $\rho_n = \nu - \angle{k}_n$, with
\begin{align}
    k_{\text{p}n} = \sum_{m=0}^\infty \p{  1 + \ec{\phi_{n,m}} }\alpha_{n,m}, \quad  \quad k_{\text{o}n} = \sum_{m=0}^\infty \alpha_{n,m} \notag
\end{align}
for the parallel and orthogonal configurations, respectively. 

From Fig. \ref{fig:xi_cdf}, we see again that as more elements are added to the \gls{ris}, the configuration obtained through the (cascaded channel) estimation procedure converges to the optimal one with an increasing probability. In fact, when $N=128$, the median power loss due to suboptimal configuration is only $\SI{-0.58}{\deci\bel}$.

\begin{remark}
    In general, the cascaded channel in \eqref{eq:cascadedChannel} is violated, as the transmitter-to-\gls{ris} and \gls{ris}-to-receiver channels are not independent but inherently correlated and dependent on applied phase-shift. Hence, multiple pilots are required to estimate the end-to-end channel even in the simplest case of a single element. As the number of elements grows, the conventional channel estimation scheme becomes more valid, and the resulting \gls{ris} configuration converges in the limit to the optimal one. 
\end{remark}

%----------------------------------------------------------------------------
% ONE-WAY COMMUNICATIONS
%----------------------------------------------------------------------------
\subsection{One-way communication}

We turn our attention now to the impact of the dependence between amplitude, phase and polarization in achieving one-way communications. Thus, consider again a single particle in isolation, with transmitter and receiver as in \eqref{eq:Direction} and polarizations as in \eqref{eq:polarization}. Denote by $A_\text{d}$ and $A_\text{u}$ the downlink (from transmitter to receiver) and uplink (from receiver to transmitter) complex responses. Then, analogously as in Section \ref{subsection:Constant_phase_shift}, for parallel polarization ($\vartheta = 0$) we have
\begin{align}
    A_\text{d} &=  \frac{ 3   }{4k}\e{-\iu \p{\rho+\frac{\pi}{2}} } \begin{bmatrix}
-u_{1, 5} - u_{5, 5} - 1 \\
u_{1, 6} + u_{5, 6} \\
u_{1, 1} + u_{5, 1} + 1
\end{bmatrix}\T \begin{bmatrix}
\ec{\varphi} \\
\es{\varphi} \\
1
\end{bmatrix}, \\
    A_\text{u} &=  \frac{ 3   }{4k}\e{-\iu \p{\rho+\frac{\pi}{2}} } \begin{bmatrix}
u_{5, 1} - u_{5, 5} - 1 \\
-u_{6, 1} + u_{6, 5} \\
u_{1, 1} - u_{1, 5} + 1
\end{bmatrix}\T \begin{bmatrix}
\ec{\varphi} \\
\es{\varphi} \\
1
\end{bmatrix}.
\end{align}

We see that, if $\varphi = 0$, i.e. transmitter and receiver are in the same angular direction, then $A_\text{d} = A_\text{u}$, and one-way communication is not possible; result that, up to some point, was expected. If, in turn, we particularize for orthogonal angular directions ($\varphi = \frac{\pi}{2}$), then 
\begin{align}
    A_\text{d} &= \frac{3}{4 k} \e{-\iu \p{\rho + \frac{\pi}{2}}} \p{1 + u_{1,1} + u_{1,6} + u_{5,1} + u_{5,6}},  \\
    A_\text{u} &= \frac{3}{4 k} \e{-\iu \p{\rho + \frac{\pi}{2}}} \p{1 + u_{1,1} - u_{6,1} - u_{1,5} + u_{6,5}}. 
\end{align}
Setting $u_{1,1} = -1$, $ u_{5,6} = \e{\iu \p{\rho + \frac{\pi}{2}}}$, and $ u_{1,6} = u_{5,1} = u_{6,1} = u_{1,5} = u_{6,5} = 0$ yields $A_\text{u} = 0$ and $A_\text{d} = 3/(4k)$, which proves that it is possible to do one-way communication under power conservation if the transmitter and receiver directions are orthogonal and polarizations are parallel.

Similarly, for orthogonal polarizations ($\vartheta = \frac{\pi}{2}$), 
\begin{equation}
\begin{aligned}
A_\text{d}&=  \frac{ 3   }{4k}\e{-\iu \p{\rho+\frac{\pi}{2}} } \begin{bmatrix}
 u_{4, 5} - u_{2, 5} \\
 u_{2, 6} - u_{4, 6} \\
 u_{2, 1} - u_{4, 1}
\end{bmatrix}\T \begin{bmatrix}
\ec{\varphi} \\
\es{\varphi} \\
1
\end{bmatrix}, \\
A_\text{u}&=  \frac{ 3   }{4k}\e{-\iu \p{\rho+\frac{\pi}{2}} } \begin{bmatrix}
 u_{5, 4} + u_{5, 2} \\
 -u_{6, 2} - u_{6, 4} \\
 u_{1, 2} + u_{1, 4}
\end{bmatrix}\T \begin{bmatrix}
\ec{\varphi} \\
\es{\varphi} \\
1
\end{bmatrix}.
\end{aligned}
\end{equation}
Setting $u_{5, 4} = -u_{5, 2}$, $u_{6, 4} = - u_{6, 2}$, $u_{1, 4} = -u_{1, 2}$, $u_{2, 1} = \e{\iu \p{\rho+\frac{\pi}{2}} }$, $u_{4, 5} = \ec{\varphi}\e{\iu \p{\rho+\frac{\pi}{2}} }$, and $u_{4, 6} = -\es{\varphi} \e{\iu \p{\rho+\frac{\pi}{2}} }$, one obtains that $A_\text{d} = 3/(2k)$ and $A_\text{u} = 0$, regardless of the angular separation $\varphi$. This implies that if the polarizations are orthogonal, one-way communications is possible for any single separation angle $\varphi$. In fact, it is possible to have one-way communication from all on-axis angles simultaneously, as is observed for the case configuration in \eqref{eq:unitaryExampleOneway} in Appendix \ref{app:2}, in which case
\begin{align}
A_\text{d}&=  \frac{3}{2 k}  \frac{1}{\sqrt{2}}, & A_\text{u}&=  0,
\end{align}
for all separation angles, $\varphi$, simultaneously.

%----------------------------------------------------------------------------
% ONE-WAY COMMUNICATIONS
%----------------------------------------------------------------------------
\subsection{Spatial multiplexing}

Finally, this section demonstrates that a \gls{ris} can support multiple orthogonal channels while meeting the reciprocity and power conservation conditions in Remarks \ref{remark:Passive} and \ref{remark:Reciprocity}. To this end, assume two transmitters and two receivers (all linearly polarized along $x$) located in the $z$-positive part of the $yz$-plane. As the \gls{ris} is in the $xy$-plane, this scenario is equivalent of a network with receivers and transmitters separated in the horizontal dimensions, as is usually the case for terrestrial networks. It is important to note that, although we focus here on proving the case of two orthogonal channels, additional channels can be supported by using different polarizations and positions off $yz$-plane. 
%\cite{I_believe_that_bjornson_actually_has_a_paper_stating_this}
Consider then the following directions and polarizations
\begin{equation}
\begin{aligned}
\hvec{p}_\text{in,t} &= \setlength\arraycolsep{3pt}\begin{bmatrix}
1 & 0 & 0
\end{bmatrix}\T, & \hvec{p}_\text{out,t} &= \setlength\arraycolsep{3pt} \begin{bmatrix}
1 & 0 & 0
\end{bmatrix}\T, \\
\hvec{r}_\text{in,t} &= \begin{bmatrix}
0 \\ -\es{\varphi_t} \\ -\ec{\varphi_t}
\end{bmatrix}, & \hvec{r}_\text{out,t} &= \begin{bmatrix}
0 \\ \es{\varphi_t-\Delta_t} \\ \ec{\varphi_t-\Delta_t}
\end{bmatrix}.
\end{aligned}
\end{equation}
where $t = \{1,2\}$ refer to the first and second end-to-end channels, i.e., the first or second pair of transmitter-receiver we aim to multiplex. Therefore, the effective scattering function between channels $t$ and $s$ is given by \eqref{eq:opt1Variable} as
\begin{equation}
    A_\text{ts} =  \hvec{p}_\text{out,t}\H \mat{S}\p{\hvec{r}_\text{out,t}, \hvec{r}_\text{in,s}} \hvec{p}_\text{in,s}
\end{equation}
with $t,s=\{1,2\}$. To achieve mutiplexing, we require the \gls{ris} to cancel the inter-channel interference while maximizing the scattering through the intended user. Mathematically, we need thus that $A_{12} = A_{21} = 0$ while maximizing $A_{11}$ and $A_{22}$. Two examples meeting these constraints are given next, where for simplicity a single \gls{ris} element is considered:
\subsubsection{$\varphi_1 = -\varphi_2 = \frac{\pi}{4}$, $\Delta_1=\Delta_2=0$} Here, it can be proved that orthogonality is achieved for $u_{6,6} = 2\sqrt{2} u_{1,5}- 2 u_{1,1} + u_{5,5} - 2$, while setting $u_{1,1} = -1$ and $u_{1,5} = 0$ ensures that power is conserved and $u_{5,5} = -\e{\iu \frac{\pi}{2}}$ maximizes the scattering of the signals under the previous conditions. The full matrix is given in \eqref{eq:unitaryExampleOrthogonalOne} in Appendix \ref{app:2}. For this particle configuration, we get $A_{11} = A_{22} = 3(1+\iu)/(4k)$ and $A_{11} = A_{22} =0$.

\subsubsection{$\varphi_1 = -\varphi_2 = \frac{\pi}{6}$, $\Delta_1=\Delta_2=\frac{\pi}{2}$} In this case, orthogonality require $u_{5,6}= - \frac{(1 + \sqrt{3})}{2} u_{1, 6}$ and 
\begin{align}
    u_{1,1} = \frac{1}{4}\p{\p{2u_{1, 5} + u_{5, 5} + u_{6, 6} + 2}\sqrt{3} + 2 u_{1, 5} - 4}.
\end{align}
Under these conditions, the scattering of the desired signal is maximized if $u_{6,6}=1$. An example of an unitary matrix meeting the previous conditions is given in \eqref{eq:unitaryExampleOrthogonalTwo}, yielding $A_{21}=A_{12} = 0$ and
\begin{align}
    A_\text{11} =  A_\text{22} =  \frac{- \iu 3\sqrt{3} }{4 k}.
\end{align}

\section{Maximizing RIS Radar Cross Section}\label{sec:optimization}
This section treats optimization of the effective \gls{rcs} $\sigma = 4 \pi \abs{  \hvec{p}_\text{out}\H\mat{S}\p{\hvec{r}_\text{out} ,\hvec{r}_\text{in}}  \hvec{p}_\text{in}  }^2$ in \eqref{eq:effRCS} given known ingoing and outgoing wave normals and polarizations $\hvec{r}_\text{in}$, $\hvec{p}_\text{in}$, $\hvec{r}_\text{out}$, and $\hvec{p}_\text{out}$. That is, we aim to maximize the \gls{ris} utility in a generic communications problem, leading to the formulation:%The problem to solve is formulated as follows;
\begin{maxi}
{\mat{V}}{ f\p{\mat{V}} = \abs{  \hvec{p}_\text{out}\H\mat{S}\p{\hvec{r}_\text{out} ,\hvec{r}_\text{in}}  \hvec{p}_\text{in}  }^2 }
{\label{eq:OptGeneral}}{}
\addConstraint{ \mat{V}\H\mat{V} = \mat{V}\mat{V}\H }{= \mat{I}_{6N}},
\end{maxi}
where $\mat{S}$ is given in \eqref{eq:S_normalized} and $\mat{V}$ is a block diagonal matrix of unitary matrices as defined in \eqref{eq:S_normalized}. 
Solutions to \eqref{eq:OptGeneral} are obtained through the following two methods:
\begin{enumerate}
    \item Numeric optimization over Riemannian manifold.
    \item Closed form solution in absence of interaction field.
\end{enumerate}

\subsection{Numeric optimization over Riemannian manifold}\label{subsec:manifoldOptimization}
In the general case, $\mat{V}$ is a block-diagonal unitary matrix, which implies each of the blocks themselves are unitary matrices, i.e., $\mat{U}_n\H\mat{U}_n = \mat{U}_n\mat{U}_n\H = \mat{I}_6$ $\forall$ $n$. Due to the unitary constraint, a practical approach is optimizing directly on the unitary matrix group, similarily to \cite{Abrudan2009}. Specifically, the \texttt{manopt}  library is employed \cite{manopt}. The Euclidean gradient is easily computed by applying standard techniques (the reader is gently referred to \cite{Hjrungnes2011} for basic theory on complex-valued matrix derivatives), yielding
\begin{align}
    \der{ f\p{\mat{V}} }{\mat{V}^*} =& S\left( \mat{D}\H - \frac{3\iu \pi}{k} \mat{D}\H \p{\mat{V}+\eye_{6N}}\H \mat{E}\H \mat{M}\H \right), \label{eq:gradientF}
\end{align}
where
\begin{align}
    \mat{D} & = \p{\frac{3 }{4 k}}^2\mat{T}\mat{O} \hvec{p}_\text{in}\hvec{p}_\text{in}\H\mat{O}\H\mat{T}\H \p{\mat{V} + \eye_{6N}}\H \times \notag \\
    &\quad\quad\quad\quad\quad\quad\quad\quad\quad\quad\quad\quad\mat{E}\H \mat{H}\H \mat{P}\H  \hvec{p}_\text{out}\hvec{p}\H_\text{out}  \mat{P} \mat{H} \mat{E},\notag\\
    \mat{E} &= \left[\mat{I}_{6N} - \frac{\iu 3\pi}{k}\p{\mat{V}+\mat{I}_{6N}}\mat{M}\right]^{-1},\notag
\end{align}
and $S(\cdot)$ selects the block diagonal blocks according to \eqref{eq:S_normalized}. Numerical optimization through the \texttt{manopt} library using the \texttt{unitaryfactory} manifold yields a local maximum of the objective function that approximately satisfies the constraint. A suboptimal solution satisfying the reciprocity constraints can be obtained by setting $S(\mat{V}) = \mat{V} \hadamard \eye_{6N}$, restricting $\mat{V}$ to be a diagonal matrix. The closed form solution given in sec. \ref{sec:closedFormSolution} is used as the initial point for the optimization.

\subsection{Closed form solution in absence of interaction field}\label{sec:closedFormSolution}
Due to the dependency between amplitude and phase of the scattered wave, see \eqref{eq:maxScattering}, maximization of the \gls{rcs} requires joint optimization of all particles even in the absence of the interaction field. Under the assumption of independence between scattering amplitude and phase, the particles can be optimized in isolation and a closed form solution can be obtained. Assuming independence between scattering amplitude and phase, solving \eqref{eq:OptGeneral} is equivalent to solving
\begin{maxi}
{\mat{V}}{ A = \Re{  \hvec{p}_\text{out}\H \mat{S}\p{\hvec{r}_\text{out} ,\hvec{r}_\text{in}}  \hvec{p}_\text{in}  }, }
{\label{eq:OptGeneral2}}{}
\addConstraint{ \mat{V}\H\mat{V} = \mat{V}\mat{V}\H }{= \mat{I}_{6N}}.
\end{maxi}
Given no interaction field, equivalent to setting $\mat{M} = \mat{0}_{6N\times 6N}$, \eqref{eq:OptGeneral2} is equivalent to
\begin{maxi}
{\mat{V}}{f\p{\mat{V}} = \Re{  \Tr{  \iu \mat{T} \mat{O} \hvec{p}_\text{in} \hvec{p}_\text{out}\H \mat{P} \mat{H} \mat{V}    } } }
{\label{eq:OptGeneral3}}{}
\addConstraint{ \mat{V}\H\mat{V} = \mat{V}\mat{V}\H }{= \mat{I}_{6N}}.
\end{maxi}
As $\mat{V}$ is block diagonal matrix of matrices $\mat{U}_n$, the objective function $f\p{\mat{V}}$ can be rewritten as
\begin{align}
    f\p{\mat{V}} = \sum_{n=1}^N  \Re{\Tr{\mat{A}_n \mat{U}_n}} \label{eq:objective3}
\end{align}
where $\mat{A}_n = \iu\mat{O} \hvec{p}_\text{in}\hvec{p}_\text{out}\H\mat{P} \e{-\iu k \p{\hvec{r}_\text{in} - \hvec{r}_\text{out} }\T\bveci{x}_n}$ and $\bveci{x}_n$ is the relative position of the $n$'th \gls{ris} particle. Similarly to what is done in \eqref{eq:maxUtility}, the objective function \eqref{eq:objective3} is maximized by computing the singular value decomposition of $\mat{A}_n = \mat{Y}_n \mati{\Sigma} \mat{k}_n\H$ and setting $\mat{U}_n = \mat{K}_n \mat{Y}_n\H$. The solution is therefore given as
\begin{align}
    \mat{V} = \begin{bmatrix} 
\mat{K}_1 \mat{Y}_1\H	& 0 		& \cdots 	& 0 \\
0 		& \mat{K}_2 \mat{Y}_2\H 	& \cdots 	& 0 \\
\vdots 		& \vdots 	& \ddots 	& \vdots \\
0 		& 0 		& \cdots 	& \mat{K}_N \mat{Y}_N\H  
\end{bmatrix}.
\end{align}
Restricting $\mat{V}$ to be a diagonal matrix yields
\begin{align}
    \mat{V} =  \e{ -\angle\p{  \iu \mat{T} \mat{O} \hvec{p}_\text{in} \hvec{p}_\text{out}\H \mat{P} \mat{H} } }  \hadamard \eye_{6N}   .
\end{align}
where $\e{\p{ \cdot}}$ is the element-wise exponential, $\angle\p{\cdot}$ yields the element-wise argument.
\iffalse
\subsection{Numeric optimization restricted to diagonal $\mat{V}$}
Restricting $\mat{V}$ to be a diagonal matrix, i.e., $\mat{V} = \diag\{\e{\iu\veci{\phi}}\}$ with\footnote{$\e{\mat{X}}$ denotes the element-wise exponential function.} $\veci{\phi}\in\mathbb{R}^{6N\times 1}$, the unitary constraint is satisfied when optimizing over $\veci{\phi}$, thus leading to the unconstrained problem
\begin{maxi}
{\veci{\phi}}{A(\veci{\phi})=\Re{   \Tr{\mat{A}\mat{E}(\mat{V})\p{\mat{V}+\mat{I}_{6N}}}  } }
{\label{eq:OptDiag}}{},
\end{maxi}
The gradient of $f(\veci{\phi})$ in \eqref{eq:OptDiag} is obtained by applying standard techniques, similarly to what was done in \eqref{eq:gradientF}, yielding
\begin{align}
    \der{A(\veci{\phi})}{\veci{\phi}} = \Imag&\left\{\diag\left\{\p{\frac{-\iu 3\pi}{k}\mat{M}\mat{E}(\mat{V})\p{\mat{V}+\mat{I}_{6N}}-\mat{I}_{6N}} \right.\right.\notag \\
    &\left.\left.\times\mat{A}\mat{E}(\mat{V})\mat{V}\phantom{\frac{1}{1}}\right\}\right\}.\label{eq:GradDiag}
\end{align}
Any descent method (e.g., gradient-descent or quasi-Newton algorithms) suffices to find a local maximum solution. 
\fi
\iffalse
\input{uniform_planar_array.tex}

\subsection{Alternative problems}
Wave absorption - This one is easy. Just optimize the isolated particle of maximum wave absorption. How does the scattering look then compared to maximum scattering?
Non-reciprocity - can the link be made completely one-way?
\fi

\section{Numerical results}\label{sec:numericalResults}
\begin{figure*}
\centering
\subfloat[][Anomalous reflection.]{\includegraphics[width=261pt]{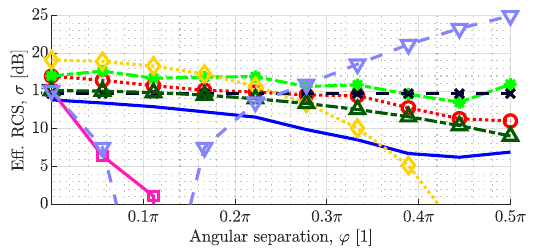}\label{fig:scenario_one}} 
\subfloat[][Specular reflection.]{\includegraphics[width=261pt]{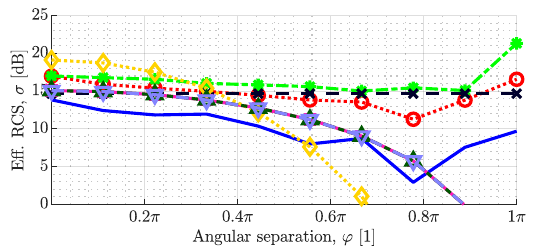}    \label{fig:scenario_two}}

\subfloat[][Constant spacing, variable elements and aperture.]{\includegraphics[width=206pt]{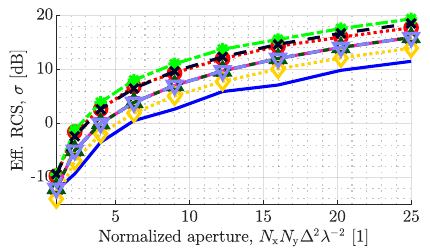}    \label{fig:scenario_three}}
\includegraphics[width=97pt]{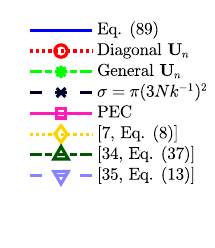}
\subfloat[][Constant elements, variable aperture and spacing.]{\includegraphics[width=206pt]{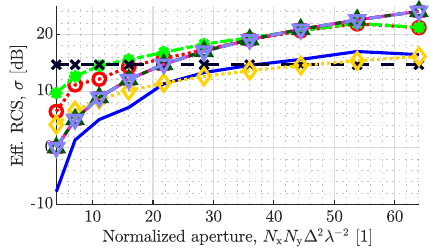}\label{fig:scenario_five}}

\caption{Simulation results for scenarios one through four.} \label{fig:sim_results}
\end{figure*}

Simulation results of the effective \gls{rcs}, $\sigma$, defined in \eqref{eq:effRCS}, are provided in Fig. \ref{fig:sim_results} for the following four scenarios (specific configurations provided in Table \ref{tab:simulation_param}): 1. Anomalous reflection, 2. Specular reflection, 3. Constant spacing, variable number of particles and aperture, and 4. Constant number of particles, variable spacing and aperture. 

The simulation results are compared with the effective \gls{rcs} obtained from an \gls{ris} model based on antenna theoretic measures \cite{tang_wireless_2021}, two different \gls{ris} models based on physical optics \cite{danufane_path-loss_2021,degli-esposti_reradiation_2022}, and the \gls{rcs} of slab of \gls{pec} of the same size.
For all scenarios, a square planar array of particles arranged in a equispaced lattice structure is simulated. The array is centered at the origin and aligned with the $xy$-plane. Unless otherwise stated, the inter-particle spacing is $\Delta = \frac{\lambda}{2}$, the number of particles along the $x$- and $y$-axis, $N_\text{x} = N_\text{y} = 8$, and $\hvec{p}_\text{in} =  \hvec{p}_\text{out} = \begin{bmatrix}
    1& 0 & 0
\end{bmatrix}\T$.
\iffalse
\begin{figure}
    \centering
    \includegraphics{figures/scattering_pattern.png}
    \caption{Scattering pattern for anomalous reflection with $\phi = \frac{\pi}{4}$.}
    \label{fig:scenario_one_pattern}
\end{figure}
\fi
Figures  \ref{fig:scenario_one} and  \ref{fig:scenario_two} show the anomalous and specular reflection scenario respectively. In both cases, it is observed that the effective \gls{rcs} remains almost constant, and does not tend to zero with the projected aperture as $\phi$ increases. In fact, the maximum effective \gls{rcs} is observed to be well approximated by $\sigma \approx \pi \p{3 N k^{-1}}^2$ regardless of the angles of illumination and reflection. This fact is in direct disagreement with the models in \cite{tang_wireless_2021,danufane_path-loss_2021,degli-esposti_reradiation_2022}, which are widely assumed in the related literature. 
\iffalse
Fig. \ref{fig:scenario_one_pattern} shows the effective \gls{rcs} obtained following sec. \ref{subsec:manifoldOptimization} for $\phi = \frac{\pi}{4}$. The gray square represents the \gls{ris} geometry and the black arrows indicates the ingoing, $\hvec{r}_\text{in}$, and outgoing, $\hvec{r}_\text{out}$, directions. The pattern shows a major lobe in the \textit{through} direction which is highlighted by the red dotted arrow. This lobe creates destructive interference with the source wave, effectively decreasing the total power of the wave in the domain behind the \gls{ris}. The energy removed by the \textit{through}-lobe is reflected towards the outgoing direction $\hvec{r}_\text{out}$. The major lobe in the \textit{through} direction serves to ensure that the total power of the system is conserved, and is therefore a requirement of power conservation.
\fi
Fig. \ref{fig:scenario_three} shows an expected linear scaling of the effective \gls{rcs} with the aperture of the \gls{ris}, which is equivalent of a scaling in the number $N = N_\text{x}N_\text{y}$ of elements squared. Finally, Fig. \ref{fig:scenario_five} shows how the effective \gls{rcs} scales as the aperture of the \gls{ris} is squeezed and stretched. Notable is the fact that with a normalized aperture of $\frac{N_\text{x}N_\text{y}}{16} = 4$, equivalent of a spacing of $\Delta = \frac{\lambda}{4}$, the maintains a \gls{rcs} of $\sigma \approx 10\text{ dB}$ while the physical optics models shows a rapid decrease in effective \gls{rcs}. As the \gls{ris} aperture is stretched out, the model shows a peak in the effective \gls{rcs} at a spacing of $\Delta \approx 0.9 \lambda$, after which the effective \gls{rcs} starts to decrease.

\begin{table}[]
    \centering
    \caption{Simulation parameters} %{\fontsize{10pt}{10pt}\selectfont
    \newsavebox\matrixbox
    \setbox\matrixbox\hbox{$\begin{bmatrix}1 \\ 0 \\ 0 \end{bmatrix}$}
    \newsavebox\matrixboxx
    \setbox\matrixboxx\hbox{$\begin{bmatrix}0 \\ -\es{\frac{\phi}{2}} \\ -\ec{\frac{\phi}{2}} \end{bmatrix}$}
    \newsavebox\matrixboxxx
    \setbox\matrixboxxx\hbox{$\begin{bmatrix}0 \\ -\es{\frac{\phi}{2}} \\ \ec{\frac{\phi}{2}} \end{bmatrix}$}
    \begin{tabular}{c | c | c | c}
    \rowcolor{gray!25}
    &&&\\[-0.85em]
    \rowcolor{gray!25}
        Scenario & $ \hvec{r}_\text{in}$& $\hvec{r}_\text{out}$ & Variable \\
        \hline
        &&&\\[-0.75em]
          \makecell{Anomalous \\ reflection} & $\begin{bmatrix}0 \\ -\es{\phi} \\ -\ec{\phi}\end{bmatrix}$  & $\begin{bmatrix} 0 \\ 0 \\ 1\end{bmatrix}$ & $\phi \in \left[0; \frac{\pi}{2}\right]$ \\
         &&&\\[-0.75em]
         \hline
         &&&\\[-0.75em]
         \makecell{Specular \\ reflection}  & $\usebox\matrixboxx$  & $\usebox\matrixboxxx$ & $\phi \in \left[0; \pi\right]$ \\
         &&&\\[-0.75em]
         \hline
         &&&\\[-0.75em]
          \makecell{Constant \\ spacing}  & $\sqrt{0.5}\begin{bmatrix}0 \\ -1 \\ -1 \end{bmatrix}$  & $\sqrt{0.5}\begin{bmatrix} 0 \\ -1 \\ 1 \end{bmatrix}$ &  $N_\text{x}, N_\text{y} \in \left[1; 10\right]$\\
         &&&\\[-0.75em]
         \hline
         &&&\\[-0.75em]
          \makecell{Constant \\ number of \\ particles}  & $\sqrt{0.5}\begin{bmatrix}0 \\ -1 \\ -1 \end{bmatrix}$  & $\sqrt{0.5}\begin{bmatrix} 0 \\ -1 \\ 1 \end{bmatrix}$ & $\Delta \in \left[\frac{\lambda}{4}; \lambda\right]$ \\
    \end{tabular} %}
    \label{tab:simulation_param}
\end{table}

%We compare to the following models: \cite{tang_wireless_2021}\\

%In paper (Wankai Tang, Ming Zheng Chen, Xiangyu Chen, Jun Yan Dai, Yu Han and Marco Di Renzo “Wireless Communications With Reconfigurable Intelligent Surface: Path Loss Modeling and Experimental Measurement,”
%IEEE Transactions on Wireless Communications, vol. 20, no. 1, pp. 421–439, Jun. 2021.) \\
%is considered a general RIS-assisted single-input single-output (SISO) wireless communication system.\\

\section{Conclusions}\label{sec:conclusion}
This work presents a narrowband analysis of single-layer metasurface based \glspl{ris}, and investigates the different wave interactions that are possible under this architecture while conserving power and ensuring \gls{ris} reciprocity.

The analysis demonstrates that all while conserving power at the element level, it is theoretically possible, using a \textit{single} \gls{ris} element, to create; a one-way communication channel, given that the in-and-outgoing directions or polarizations are orthogonal, or at least two orthogonal communication channels without utilizing orthogonal polarizations. This means that the same aperture can be utilized to serve multiple communication links without causing interference. 

The analysis also disproves the common assumption that a \gls{ris} is capable of doing constant amplitude phase-shifting for all separation angles simultaneously. It investigates the implications of this limitation for a \gls{ris} assisted communication scenarios. Due to the lack of constant amplitude phase-shifting, the cascaded channel model, $h = h_1 \phi h_2$, is not valid in multipath scenarios. This has implications for systems relying on signal-space channel estimation, as the channel does not remain deterministic as the phase-shift of an element is changed. However, the analysis also shows that given a sufficiently high number of reconfigurable \gls{ris} elements, e.g. 128, the loss in signal power due to this randomness becomes negligible.% and can be ignored.

For a planar array, the analysis finds that the maximum \gls{rcs} of a \gls{ris} is well approximated by $\sigma \approx \pi \p{3 N k^{-1}}^2$, where $N$ is the total number of elements and $\lambda$ is the wavelength. A higher frequency thus requires more elements to achieve the same effective \gls{rcs} and thereby signal power. The simulations shows the effective \gls{rcs} does not scale with the projected aperture of the \gls{ris}, as the effective \gls{rcs} remains in the same order of magnitude as the angles of incidence and reflection approaches angles parallel to the \gls{ris} plane. This is in contrast to what is predicted by the physical optics based models \cite{danufane_path-loss_2021,degli-esposti_reradiation_2022}, which predicts the effective \gls{rcs} tending to zero as the angles of incidence and reflection becomes parallel to the \gls{ris} plane.

\appendices

\section{Relations and conditions for point actors}\label{app:1}

This appendix derives conditions for passive point actors and reciprocal point actors.
The current distribution for a point actor positioned at $\vec{x}'$ is given as
\begin{align}
\begin{bmatrix}
\veci{j}\p{\vec{x}} \\ \veci{m}\p{\vec{x}}
\end{bmatrix} = \delta\p{\vec{x} - \vec{x}'} \begin{bmatrix}
\vec{j} \\ \vec{m}
\end{bmatrix}.
\end{align}
\subsection{Passive and lossless conditions}
For a point scatterer defined by the polarizability matrix $\mat{X}$, being passive implies that the supplied power $P_\text{s}\leq 0$, and being lossless implies that the supplied power $P_\text{s}\geq 0$. As such, for the point scatterer to be both passive and lossless implies that $P_\text{s} = 0$. The time-averaged power \textit{supplied} by the point actor is given as \cite[sec. 1.7.3]{balanis_advanced_2012}
\begin{align}
P_\text{s} &= \frac{-1}{2} \iiint_V \Re{ \begin{bmatrix}
    \vec{e}\p{\vec{x}} \\ \vec{h}\p{\vec{x}}
\end{bmatrix}\H \begin{bmatrix}
    \veci{j}\p{\vec{x}} \\ \veci{m}\p{\vec{x}}
\end{bmatrix}  } \dd{\vec{x}}, \label{eq:inducedPower} \\
&= \frac{-1}{2} \Re{ \begin{bmatrix}
    \vec{e}\p{\vec{x}'} \\ \vec{h}\p{\vec{x}'}
\end{bmatrix}\H  \begin{bmatrix}
\vec{j} \\ \vec{m}
\end{bmatrix}  } .
\end{align}
Applying \eqref{eq:greensMatrix4} and \eqref{eq:greensMatrix5} yields
\begin{align}
P_\text{s} &= \frac{-1}{2} \Re{ \begin{bmatrix}
\vec{e}_{\text{a}} \\ \vec{h}_{\text{a}}
\end{bmatrix}\H \p{  \mat{X}\H \mat{G}\p{\mat{0}_{3\times 1}}\H \mat{X}  + \mat{X}} \begin{bmatrix}
\vec{e}_{\text{a}} \\ \vec{h}_{\text{a}}
\end{bmatrix}  }, \notag \\
&= \frac{-1}{2} \begin{bmatrix}
    \vec{e}_\text{a} \\ \vec{h}_\text{a}
\end{bmatrix}\H \p{ \frac{\mat{X} + \mat{X}\H}{2} +\mat{X}\H \mat{G}_0 \mat{X}}  \begin{bmatrix}
    \vec{e}_\text{a} \\ \vec{h}_\text{a}
\end{bmatrix}.
\end{align}
where $\begin{bmatrix}
    \vec{e}_\text{a} & \vec{h}_\text{a}
\end{bmatrix}\T$ is the field acting upon the point scatterer and $\mat{G}_0$ is given by \eqref{eq:G0}. The scatterer is lossless and passive, $P_\text{s} = 0$, i.e., if and only if \eqref{eq:PassiveCondition} holds.

\subsection{Reciprocity conditions}

To investigate channel reciprocity, we consider a single point scatterer and two point actors positioned at spatial coordinates $\vec{x}_0$, $\vec{x}_1$, and $\vec{x}_2$ respectively. The first point actor carries currents $\vec{j}$ and $\vec{m}$ which radiates an electromagnetic field. The field is measured by the other actor along the dimension dictated by the measurement vector $\vec{a}$, yielding the signal $h_1$. $\vec{a}$ corresponds to an effective gain pattern in conventional antenna theory terms. $\vec{a}$ is split into its electric and magnetic parts as $\vec{a} = \begin{bmatrix} \vec{a}_\text{e}\T & \vec{a}_\text{m}\T \end{bmatrix}\T$. Afterwards, the positions are switched and the sign of the effective magnetic currents $\vec{m}$ and the magnetic measurement vector $\vec{a}_\text{m}$ is switched to reverse the direction of radiation and measurement. Repeating the process, yields the measured signal $h_2$. $h_1$ and $h_2$ are given as
\begin{align}
h_1 &= \begin{bmatrix} \vec{a}_\text{e} \\ \vec{a}_\text{m} \end{bmatrix}\T \big(\mat{G}\p{\vec{x}_2 - \vec{x}_1} \notag \\ 
&+  \mat{G}\p{\vec{x}_2 - \vec{x}_0} \mat{X}   \mat{G}\p{\vec{x}_0 - \vec{x}_1}  \big) \begin{bmatrix} \vec{j} \\ \vec{m} \end{bmatrix}, \\
h_2 &= \begin{bmatrix} \vec{a}_\text{e} \\ -\vec{a}_\text{m} \end{bmatrix}\T \big(\mat{G}\p{\vec{x}_1 - \vec{x}_2} \notag \\  
&+  \mat{G}\p{\vec{x}_1 - \vec{x}_0} \mat{X}   \mat{G}\p{\vec{x}_0 - \vec{x}_2}  \big) \begin{bmatrix} \vec{j} \\ -\vec{m} \end{bmatrix}.
\end{align}
For channel reciprocity, we require that the two measured signals are equal, i.e., $h_1 = h_2$. Defining the submatrices of $\mat{X}$ according to \eqref{eq:Xsubmatrices} and exploiting the symmetry of the free space Green's matrix, the conditions for reciprocity yield; $\mat{X}_\text{ee} = \mat{X}_\text{ee}\T$, $\mat{X}_\text{mm} = \mat{X}_\text{mm}\T$, and $\mat{X}_\text{em} = -\mat{X}_\text{me}\T$.

\section{Unitary matrices for examples}\label{app:2}
For all example matrices, $u_{n,m} = \begin{cases}
    0 & n\neq m\\
    -1 & n = m
\end{cases}$ unless otherwise stated.

\begin{enumerate}[leftmargin=14pt]
    \item Constant Amplitude phase-shifting for $\varphi=0$ and $\vartheta = 0$:
    \begin{align}
   u_{1,1} = -u_{5,5} = \e{\iu \p{\rho + \frac{\pi}{2}}}. \label{eq:unitaryExampleBroadsideParallel}
\end{align}
    \item Constant Amplitude phase-shifting for $\varphi=\frac{\pi}{2}$ and $\vartheta = 0$:
    \begin{align}
    u_{1,1} = \e{\iu \p{\rho + \frac{\pi}{2}}},\,\, u_{5,5} = 0,\,\, u_{6,5} = u_{5,6} = -1.\label{eq:unitaryExampleOrthogonalParallel}
\end{align}
    \item Constant amplitude phase-shifting for all on-axis angles, orthogonal polarization:
    \begin{equation}
    \begin{aligned}
u_{1,1} &= 0,\,\, u_{2,1} = u_{1,2} = -u_{4,1} = u_{4,1} = \tfrac{\sqrt{2}}{2} \e{\iu \p{\rho + \frac{\pi}{2}}},\\
u_{2,2} &= u_{4,2} = -u_{4,4} = -u_{2,4} = \tfrac{-1}{2}. \label{eq:unitaryExampleOrthogonalPolarization}
\end{aligned}
\end{equation}
    \item One-way communications for all on-axis angles, orthogonal polarization.
    \begin{equation}
    \begin{aligned}
u_{1,1} &= u_{3,3} = u_{4,4} = 0,\,\, u_{2,1} = -u_{4,1} = \tfrac{\sqrt{2}}{2} \e{\iu \p{\rho + \frac{\pi}{2}}},  \\
u_{1,3} &= u_{3,4} = -1,\,\, u_{2,2} = u_{4,2} =  \tfrac{-\sqrt{2}}{2}. \label{eq:unitaryExampleOneway}
\end{aligned}
\end{equation}
    \item Spatial multiplexing, parallel receiver and transmitter:
    \begin{align}
u_{5,5} = u_{6,6} = -\iu. \label{eq:unitaryExampleOrthogonalOne}
\end{align}
    \item Spatial multiplexing, orthogonal receiver and transmitter:
    \begin{equation}
    \begin{aligned}
u_{6,6} &= 1,\,\, u_{1,1} = -u_{2,2} = -1 + \tfrac{\sqrt{3}}{2},\\
u_{2,1} &= u_{1,2} = \tfrac{\sqrt{-3 + 4 \sqrt{3}}}{2}. \label{eq:unitaryExampleOrthogonalTwo}
\end{aligned}
\end{equation}
\end{enumerate}

\bibliographystyle{setup/IEEEtran}
\bibliography{setup/IEEEabrv,setup/references}

\end{document}